\catcode`@=11 
%
%
%

\font\fourteenrm=cmr10 scaled\magstep2
\font\twelverm=cmr10 scaled\magstep1
\font\ninerm=cmr9	     \font\sixrm=cmr6

\font\fourteenbf=cmbx10 scaled\magstep2
\font\twelvebf=cmbx10 scaled\magstep1
\font\ninebf=cmbx9	      \font\sixbf=cmbx6
\font\seventeeni=cmmi10 scaled\magstep3	    \skewchar\seventeeni='177
\font\fourteeni=cmmi10 scaled\magstep2	    \skewchar\fourteeni='177
\font\twelvei=cmmi10 scaled\magstep1	    \skewchar\twelvei='177
\font\ninei=cmmi9			    \skewchar\ninei='177
\font\sixi=cmmi6			    \skewchar\sixi='177
\font\seventeensy=cmsy10 scaled\magstep3    \skewchar\seventeensy='60
\font\fourteensy=cmsy10 scaled\magstep2	    \skewchar\fourteensy='60
\font\twelvesy=cmsy10 scaled\magstep1	    \skewchar\twelvesy='60
\font\ninesy=cmsy9			    \skewchar\ninesy='60
\font\sixsy=cmsy6			    \skewchar\sixsy='60

\font\fourteenex=cmex10 scaled\magstep2
\font\twelveex=cmex10 scaled\magstep1

\font\fourteensl=cmsl10 scaled\magstep2
\font\twelvesl=cmsl10 scaled\magstep1
\font\ninesl=cmsl9

\font\fourteenit=cmti10 scaled\magstep2
\font\twelveit=cmti10 scaled\magstep1
\font\twelvett=cmtt10 scaled\magstep1
\font\twelvecp=cmcsc10 scaled\magstep1
\font\tencp=cmcsc10
\newfam\cpfam
%
%
\newcount\f@ntkey	     \f@ntkey=0
\def\samef@nt{\relax \ifcase\f@ntkey \rm \or\oldstyle \or\or
	 \or\it \or\sl \or\bf \or\tt \or\caps \fi }
\def\fourteenpoint{\relax
    \textfont0=\fourteenrm	    \scriptfont0=\tenrm
    \scriptscriptfont0=\sevenrm
     \def\rm{\fam0 \fourteenrm \f@ntkey=0 }\relax
    \textfont1=\fourteeni	    \scriptfont1=\teni
    \scriptscriptfont1=\seveni
     \def\oldstyle{\fam1 \fourteeni\f@ntkey=1 }\relax
    \textfont2=\fourteensy	    \scriptfont2=\tensy
    \scriptscriptfont2=\sevensy
    \textfont3=\fourteenex     \scriptfont3=\fourteenex
    \scriptscriptfont3=\fourteenex
    \def\it{\fam\itfam \fourteenit\f@ntkey=4 }\textfont\itfam=\fourteenit
    \def\sl{\fam\slfam \fourteensl\f@ntkey=5 }\textfont\slfam=\fourteensl
    \scriptfont\slfam=\tensl
    \def\bf{\fam\bffam \fourteenbf\f@ntkey=6 }\textfont\bffam=\fourteenbf
    \scriptfont\bffam=\tenbf	 \scriptscriptfont\bffam=\sevenbf
    \def\tt{\fam\ttfam \twelvett \f@ntkey=7 }\textfont\ttfam=\twelvett
    \h@big=11.9\p@ \h@Big=16.1\p@ \h@bigg=20.3\p@ \h@Bigg=24.5\p@
    \def\caps{\fam\cpfam \twelvecp \f@ntkey=8 }\textfont\cpfam=\twelvecp
    \setbox\strutbox=\hbox{\vrule height 12pt depth 5pt width\z@}\relax
    \samef@nt}
\def\twelvepoint{\relax
    \textfont0=\twelverm	  \scriptfont0=\ninerm
    \scriptscriptfont0=\sevenrm
     \def\rm{\fam0 \twelverm \f@ntkey=0 }\relax
    \textfont1=\twelvei		  \scriptfont1=\ninei
    \scriptscriptfont1=\seveni
     \def\oldstyle{\fam1 \twelvei\f@ntkey=1 }\relax
    \textfont2=\twelvesy	  \scriptfont2=\ninesy
    \scriptscriptfont2=\sevensy
    \textfont3=\twelveex	  \scriptfont3=\twelveex
    \scriptscriptfont3=\twelveex
    \def\it{\fam\itfam \twelveit \f@ntkey=4 }\textfont\itfam=\twelveit
    \def\sl{\fam\slfam \twelvesl \f@ntkey=5 }\textfont\slfam=\twelvesl
    \scriptfont\slfam=\ninesl
    \def\bf{\fam\bffam \twelvebf \f@ntkey=6 }\textfont\bffam=\twelvebf
    \scriptfont\bffam=\ninebf	  \scriptscriptfont\bffam=\sevenbf
    \def\tt{\fam\ttfam \twelvett \f@ntkey=7 }\textfont\ttfam=\twelvett
    \h@big=10.2\p@ \h@Big=13.8\p@ \h@bigg=17.4\p@ \h@Bigg=21.0\p@
    \def\caps{\fam\cpfam \twelvecp \f@ntkey=8 }\textfont\cpfam=\twelvecp
    \setbox\strutbox=\hbox{\vrule height 10pt depth 4pt width\z@}\relax
    \samef@nt}
\def\tenpoint{\relax
    \textfont0=\tenrm	       \scriptfont0=\sevenrm
    \scriptscriptfont0=\fiverm
    \def\rm{\fam0 \tenrm \f@ntkey=0 }\relax
    \textfont1=\teni	       \scriptfont1=\seveni
    \scriptscriptfont1=\fivei
    \def\oldstyle{\fam1 \teni \f@ntkey=1 }\relax
    \textfont2=\tensy	       \scriptfont2=\sevensy
    \scriptscriptfont2=\fivesy
    \textfont3=\tenex	       \scriptfont3=\tenex
    \scriptscriptfont3=\tenex
    \def\it{\fam\itfam \tenit \f@ntkey=4 }\textfont\itfam=\tenit
    \def\sl{\fam\slfam \tensl \f@ntkey=5 }\textfont\slfam=\tensl
    \def\bf{\fam\bffam \tenbf \f@ntkey=6 }\textfont\bffam=\tenbf
    \scriptfont\bffam=\sevenbf	   \scriptscriptfont\bffam=\fivebf
    \def\tt{\fam\ttfam \tentt \f@ntkey=7 }\textfont\ttfam=\tentt
    \def\caps{\fam\cpfam \tencp \f@ntkey=8 }\textfont\cpfam=\tencp
    \h@big=8.5\p@ \h@Big=11.5\p@ \h@bigg=14.5\p@ \h@Bigg=17.5\p@
    \setbox\strutbox=\hbox{\vrule height 8.5pt depth 3.5pt width\z@}\relax
    \samef@nt}
%
%
%
%
\newdimen\h@big  \h@big=8.5\p@
\newdimen\h@Big  \h@Big=11.5\p@
\newdimen\h@bigg  \h@bigg=14.5\p@
\newdimen\h@Bigg  \h@Bigg=17.5\p@
\def\big#1{{\hbox{$\left#1\vbox to\h@big{}\right.\n@space$}}}
\def\Big#1{{\hbox{$\left#1\vbox to\h@Big{}\right.\n@space$}}}
\def\bigg#1{{\hbox{$\left#1\vbox to\h@bigg{}\right.\n@space$}}}
\def\Bigg#1{{\hbox{$\left#1\vbox to\h@Bigg{}\right.\n@space$}}}
%
%
%
\normalbaselineskip = 20pt plus 0.2pt minus 0.1pt
\normallineskip = 1.5pt plus 0.1pt minus 0.1pt
\normallineskiplimit = 1.5pt
\newskip\normaldisplayskip
\normaldisplayskip = 20pt plus 5pt minus 10pt
\newskip\normaldispshortskip
\normaldispshortskip = 6pt plus 5pt
\newskip\normalparskip
\normalparskip = 6pt plus 2pt minus 1pt
\newskip\skipregister
\skipregister = 5pt plus 2pt minus 1.5pt
\newif\ifsingl@	   \newif\ifdoubl@
\newif\iftwelv@	   \twelv@true
\def\singlespace{\singl@true\doubl@false\spaces@t}
\def\doublespace{\singl@false\doubl@true\spaces@t}
\def\normalspace{\singl@false\doubl@false\spaces@t}
\def\Tenpoint{\tenpoint\twelv@false\spaces@t}
\def\Twelvepoint{\twelvepoint\twelv@true\spaces@t}
\def\spaces@t{\relax
 \iftwelv@\ifsingl@\subspaces@t3:4;\else\subspaces@t1:1;\fi
 \else\ifsingl@\subspaces@t3:5;\else\subspaces@t4:5;\fi\fi
 \ifdoubl@\multiply\baselineskip by 5 \divide\baselineskip by 4 \fi}
\def\subspaces@t#1:#2;{\baselineskip=\normalbaselineskip
\multiply\baselineskip by #1\divide\baselineskip by #2%
\lineskip = \normallineskip
\multiply\lineskip by #1\divide\lineskip by #2%
\lineskiplimit = \normallineskiplimit
\multiply\lineskiplimit by #1\divide\lineskiplimit by #2%
\parskip = \normalparskip
\multiply\parskip by #1\divide\parskip by #2%
\abovedisplayskip = \normaldisplayskip
\multiply\abovedisplayskip by #1\divide\abovedisplayskip by #2%
\belowdisplayskip = \abovedisplayskip
\abovedisplayshortskip = \normaldispshortskip
\multiply\abovedisplayshortskip by #1%
\divide\abovedisplayshortskip by #2%
\belowdisplayshortskip = \abovedisplayshortskip
\advance\belowdisplayshortskip by \belowdisplayskip
\divide\belowdisplayshortskip by 2
\smallskipamount = \skipregister
\multiply\smallskipamount by #1\divide\smallskipamount by #2%
\medskipamount = \smallskipamount \multiply\medskipamount by 2
\bigskipamount = \smallskipamount \multiply\bigskipamount by 4 }
\def\normalbaselines{ \baselineskip=\normalbaselineskip%
   \lineskip=\normallineskip \lineskiplimit=\normallineskip%
   \iftwelv@\else \multiply\baselineskip by 4 \divide\baselineskip by 5%
     \multiply\lineskiplimit by 4 \divide\lineskiplimit by 5%
     \multiply\lineskip by 4 \divide\lineskip by 5 \fi }
\Twelvepoint  
\interlinepenalty=50
\interfootnotelinepenalty=5000
\predisplaypenalty=9000
\postdisplaypenalty=500
\hfuzz=1pt
\vfuzz=0.2pt
%
%
%
\def\pagecontents{%
   \ifvoid\topins\else\unvbox\topins\vskip\skip\topins\fi
   \dimen@ = \dp255 \unvbox255
   \ifvoid\footins\else\vskip\skip\footins\footrule\unvbox\footins\fi
   \ifr@ggedbottom \kern-\dimen@ \vfil \fi }
\def\makeheadline{\vbox to 0pt{ \skip@=\topskip
      \advance\skip@ by -12pt \advance\skip@ by -2\normalbaselineskip
      \vskip\skip@ \line{\vbox to 12pt{}\the\headline} \vss
      }\nointerlineskip}
\def\makefootline{\baselineskip = 1.5\normalbaselineskip
		 \line{\the\footline}}
\newif\iffrontpage
\newif\ifletterstyle
\newif\ifp@genum
\def\nopagenumbers{\p@genumfalse}
\def\pagenumbers{\p@genumtrue}
\pagenumbers
\newtoks\paperheadline
\newtoks\letterheadline
\newtoks\letterfrontheadline
\newtoks\lettermainheadline
\newtoks\paperfootline
\newtoks\letterfootline
\newtoks\date
\newtoks\Month
\footline={\ifletterstyle\the\letterfootline\else\the\paperfootline\fi}
\paperfootline={\hss\iffrontpage\else\ifp@genum\tenrm\folio\hss\fi\fi}
\letterfootline={\hfil}
\headline={\ifletterstyle\the\letterheadline\else\the\paperheadline\fi}
\paperheadline={\hfil}
\letterheadline{\iffrontpage\the\letterfrontheadline
     \else\the\lettermainheadline\fi}
\lettermainheadline={\rm\ifp@genum page \ \folio\fi\hfil\the\date}
\def\monthname{\relax\ifcase\month 0/\or January\or February\or
   March\or April\or May\or June\or July\or August\or September\or
   October\or November\or December\else\number\month/\fi}
\date={\monthname\ \number\day, \number\year}
\Month={\monthname\ \number\year}
\countdef\pagenumber=1  \pagenumber=1
\def\advancepageno{\global\advance\pageno by 1
   \ifnum\pagenumber<0 \global\advance\pagenumber by -1
    \else\global\advance\pagenumber by 1 \fi \global\frontpagefalse }
\def\folio{\ifnum\pagenumber<0 \romannumeral-\pagenumber
	   \else \number\pagenumber \fi }
\def\footrule{\dimen@=\prevdepth\nointerlineskip
   \vbox to 0pt{\vskip -0.25\baselineskip \hrule width 0.35\hsize \vss}
   \prevdepth=\dimen@ }
\newtoks\foottokens
\foottokens={\Tenpoint\singlespace}
\newdimen\footindent
\footindent=24pt
\def\vfootnote#1{\insert\footins\bgroup  \the\foottokens
   \interlinepenalty=\interfootnotelinepenalty \floatingpenalty=20000
   \splittopskip=\ht\strutbox \boxmaxdepth=\dp\strutbox
   \leftskip=\footindent \rightskip=\z@skip
   \parindent=0.5\footindent \parfillskip=0pt plus 1fil
   \spaceskip=\z@skip \xspaceskip=\z@skip
   \Textindent{$ #1 $}\footstrut\futurelet\next\fo@t}
\def\Textindent#1{\noindent\llap{#1\enspace}\ignorespaces}
\def\footnote#1{\attach{#1}\vfootnote{#1}}

\let\footsymbol=\star
\newcount\lastf@@t	     \lastf@@t=-1
\newcount\footsymbolcount    \footsymbolcount=0
\newif\ifPhysRev
\def\footsymbolgen{\relax \ifPhysRev \iffrontpage \NPsymbolgen\else
      \PRsymbolgen\fi \else \NPsymbolgen\fi
   \global\lastf@@t=\pageno \footsymbol }
\def\NPsymbolgen{\ifnum\footsymbolcount<0 \global\footsymbolcount=0\fi
   {\iffrontpage \else \advance\lastf@@t by 1 \fi
    \ifnum\lastf@@t<\pageno \global\footsymbolcount=0
     \else \global\advance\footsymbolcount by 1 \fi }
   \ifcase\footsymbolcount \fd@f\star\or \fd@f\dagger\or \fd@f\ast\or
    \fd@f\ddagger\or \fd@f\natural\or \fd@f\diamond\or \fd@f\bullet\or
    \fd@f\nabla\else \fd@f\dagger\global\footsymbolcount=0 \fi }
\def\fd@f#1{\xdef\footsymbol{#1}}
\def\PRsymbolgen{\ifnum\footsymbolcount>0 \global\footsymbolcount=0\fi
      \global\advance\footsymbolcount by -1
      \xdef\footsymbol{\sharp\number-\footsymbolcount} }
\def\space@ver#1{\let\@sf=\empty \ifmmode #1\else \ifhmode
   \edef\@sf{\spacefactor=\the\spacefactor}\unskip${}#1$\relax\fi\fi}
\def\attach#1{\space@ver{\strut^{\mkern 2mu #1} }\@sf\ }
%
%
%
\newcount\chapternumber	     \chapternumber=0
\newcount\sectionnumber	     \sectionnumber=0
\newcount\equanumber	     \equanumber=0
\let\chapterlabel=\relax
\newtoks\chapterstyle	     \chapterstyle={\Number}
\newskip\chapterskip	     \chapterskip=\bigskipamount
\newskip\sectionskip	     \sectionskip=\medskipamount
\newskip\headskip	     \headskip=8pt plus 3pt minus 3pt
\newdimen\chapterminspace    \chapterminspace=15pc
\newdimen\sectionminspace    \sectionminspace=10pc
\newdimen\referenceminspace  \referenceminspace=25pc
\def\chapterreset{\global\advance\chapternumber by 1
   \ifnum\equanumber<0 \else\global\equanumber=0\fi
   \sectionnumber=0 \makel@bel}
\def\makel@bel{\xdef\chapterlabel{%
\the\chapterstyle{\the\chapternumber}.}}
\def\sectionlabel{\number\sectionnumber \quad }
\def\alphabetic#1{\count255='140 \advance\count255 by #1\char\count255}
\def\Alphabetic#1{\count255='100 \advance\count255 by #1\char\count255}
\def\Roman#1{\uppercase\expandafter{\romannumeral #1}}
\def\roman#1{\romannumeral #1}
\def\Number#1{\number #1}
\def\unnumberedchapters{\let\makel@bel=\relax \let\chapterlabel=\relax
\let\sectionlabel=\relax \equanumber=-1 }
\def\titlestyle#1{\par\begingroup \interlinepenalty=9999
     \leftskip=0.02\hsize plus 0.23\hsize minus 0.02\hsize
     \rightskip=\leftskip \parfillskip=0pt
     \hyphenpenalty=9000 \exhyphenpenalty=9000
     \tolerance=9999 \pretolerance=9000
     \spaceskip=0.333em \xspaceskip=0.5em
     \iftwelv@\fourteenpoint\else\twelvepoint\fi
   \noindent #1\par\endgroup }
\def\spacecheck#1{\dimen@=\pagegoal\advance\dimen@ by -\pagetotal
   \ifdim\dimen@<#1 \ifdim\dimen@>0pt \vfil\break \fi\fi}
\def\chapter#1{\par \penalty-300 \vskip\chapterskip
   \spacecheck\chapterminspace
   \chapterreset \titlestyle{\chapterlabel \ #1}
   \nobreak\vskip\headskip \penalty 30000
   \wlog{\string\chapter\ \chapterlabel} }

\def\section#1{\par \ifnum\the\lastpenalty=30000\else
   \penalty-200\vskip\sectionskip \spacecheck\sectionminspace\fi
   \wlog{\string\section\ \chapterlabel \the\sectionnumber}
   \global\advance\sectionnumber by 1  \noindent
   {\caps\enspace\chapterlabel \sectionlabel #1}\par
   \nobreak\vskip\headskip \penalty 30000 }
\def\subsection#1{\par
   \ifnum\the\lastpenalty=30000\else \penalty-100\smallskip \fi
   \noindent\undertext{#1}\enspace \vadjust{\penalty5000}}

\def\undertext#1{\vtop{\hbox{#1}\kern 1pt \hrule}}
\def\ack{\par\penalty-100\medskip \spacecheck\sectionminspace
   \line{\fourteenrm\hfil ACKNOWLEDGMENTS\hfil}\nobreak\vskip\headskip }
%
\def\APPENDIX#1#2{\par\penalty-300\vskip\chapterskip
   \spacecheck\chapterminspace \chapterreset \xdef\chapterlabel{#1}
   \titlestyle{APPENDIX #2} \nobreak\vskip\headskip \penalty 30000
   \wlog{\string\Appendix\ \chapterlabel} }
\def\Appendix#1{\APPENDIX{#1}{#1}}
\def\appendix{\APPENDIX{A}{}}
%
%
%
\def\eqname#1{\relax \ifnum\equanumber<0
     \xdef#1{{\rm(\number-\equanumber)}}\global\advance\equanumber by -1
    \else \global\advance\equanumber by 1
      \xdef#1{{\rm(\chapterlabel \number\equanumber)}} \fi}
\def\eq{\eqname\?\?}

\def\eqinsert#1{\noalign{\dimen@=\prevdepth \nointerlineskip
   \setbox0=\hbox to\displaywidth{\hfil #1}
   \vbox to 0pt{\vss\hbox{$\!\box0\!$}\kern-0.5\baselineskip}
   \prevdepth=\dimen@}}
%

%

%

%
%
\def\GENITEM#1;#2{\par \hangafter=0 \hangindent=#1
    \Textindent{$ #2 $}\ignorespaces}
\outer\def\newitem#1=#2;{\gdef#1{\GENITEM #2;}}
\newdimen\itemsize		  \itemsize=30pt
\newitem\item=1\itemsize;
\newitem\sitem=1.75\itemsize;	  
\newitem\ssitem=2.5\itemsize;	  
\outer\def\newlist#1=#2&#3&#4;{\toks0={#2}\toks1={#3}%
   \count255=\escapechar \escapechar=-1
   \alloc@0\list\countdef\insc@unt\listcount	 \listcount=0
   \edef#1{\par
      \countdef\listcount=\the\allocationnumber
      \advance\listcount by 1
      \hangafter=0 \hangindent=#4
      \Textindent{\the\toks0{\listcount}\the\toks1}}
   \expandafter\expandafter\expandafter
    \edef\c@t#1{begin}{\par
      \countdef\listcount=\the\allocationnumber \listcount=1
      \hangafter=0 \hangindent=#4
      \Textindent{\the\toks0{\listcount}\the\toks1}}
   \expandafter\expandafter\expandafter
    \edef\c@t#1{con}{\par \hangafter=0 \hangindent=#4 \noindent}
   \escapechar=\count255}
\def\c@t#1#2{\csname\string#1#2\endcsname}
\newlist\point=\Number&.&1.0\itemsize;
\newlist\subpoint=(\alphabetic&)&1.75\itemsize;
\newlist\subsubpoint=(\roman&)&2.5\itemsize;
%

%
%
%
\newcount\referencecount     \referencecount=0
\newif\ifreferenceopen	     \newwrite\referencewrite
\newtoks\rw@toks
\def\NPrefmark#1{\attach{\scriptscriptstyle [ #1 ] }}
\let\PRrefmark=\attach
\def\refmark#1{\relax\ifPhysRev\PRrefmark{#1}\else\NPrefmark{#1}\fi}
\def\refend{\refmark{\number\referencecount}}
\newcount\lastrefsbegincount \lastrefsbegincount=0
\def\refsend{\refmark{\count255=\referencecount
   \advance\count255 by-\lastrefsbegincount
   \ifcase\count255 \number\referencecount
   \or \number\lastrefsbegincount,\number\referencecount
   \else \number\lastrefsbegincount-\number\referencecount \fi}}
\def\refch@ck{\chardef\rw@write=\referencewrite
   \ifreferenceopen \else \referenceopentrue
   \immediate\openout\referencewrite=reference.aux \fi}
%
{\catcode`\^^M=\active 
  \gdef\obeyendofline{\catcode`\^^M\active \let^^M\ }}%
%
{\catcode`\^^M=\active 
  \gdef\ignoreendofline{\catcode`\^^M=5}}
{\obeyendofline\gdef\rw@start#1{\def\t@st{#1} \ifx\t@st\blankend%
\endgroup \@sf \relax \else \ifx\t@st\bl@nkend \endgroup \@sf \relax%
\else \rw@begin#1
\backtotext
\fi \fi } }
{\obeyendofline\gdef\rw@begin#1
{\def\n@xt{#1}\rw@toks={#1}\relax%
\rw@next}}
\def\blankend{}
{\obeylines\gdef\bl@nkend{
}}
\newif\iffirstrefline  \firstreflinetrue
\def\rwr@teswitch{\ifx\n@xt\blankend \let\n@xt=\rw@begin %
 \else\iffirstrefline \global\firstreflinefalse%
\immediate\write\rw@write{\noexpand\obeyendofline \the\rw@toks}%
\let\n@xt=\rw@begin%
      \else\ifx\n@xt\rw@@d \def\n@xt{\immediate\write\rw@write{%
	\noexpand\ignoreendofline}\endgroup \@sf}%
	     \else \immediate\write\rw@write{\the\rw@toks}%
	     \let\n@xt=\rw@begin\fi\fi \fi}
\def\rw@next{\rwr@teswitch\n@xt}
\def\rw@@d{\backtotext} \let\rw@end=\relax
\let\backtotext=\relax

\newdimen\refindent	\refindent=30pt
\def\refitem#1{\par \hangafter=0 \hangindent=\refindent \Textindent{#1}}
\def\REFNUM#1{\space@ver{}\refch@ck \firstreflinetrue%
 \global\advance\referencecount by 1 \xdef#1{\the\referencecount}}
\def\refnum#1{\space@ver{}\refch@ck \firstreflinetrue%
 \global\advance\referencecount by 1 \xdef#1{\the\referencecount}\refend}
\def\REF#1{\REFNUM#1%
 \immediate\write\referencewrite{%
 \noexpand\refitem{#1.}}%
\begingroup\obeyendofline\rw@start}
\def\ref{\refnum\?%
 \immediate\write\referencewrite{\noexpand\refitem{\?.}}%
\begingroup\obeyendofline\rw@start}
\def\Ref#1{\refnum#1%
 \immediate\write\referencewrite{\noexpand\refitem{#1.}}%
\begingroup\obeyendofline\rw@start}
\def\REFS#1{\REFNUM#1\global\lastrefsbegincount=\referencecount
\immediate\write\referencewrite{\noexpand\refitem{#1.}}%
\begingroup\obeyendofline\rw@start}
\newcount\figurecount	  \figurecount=0
\newif\iffigureopen	  \newwrite\figurewrite
\def\figch@ck{\chardef\rw@write=\figurewrite \iffigureopen\else
   \immediate\openout\figurewrite=figures.aux
   \figureopentrue\fi}
\def\FIGNUM#1{\space@ver{}\figch@ck \firstreflinetrue%
 \global\advance\figurecount by 1 \xdef#1{\the\figurecount}}
\def\FIG#1{\FIGNUM#1
   \immediate\write\figurewrite{\noexpand\refitem{#1.}}%
   \begingroup\obeyendofline\rw@start}
\def\FIGFIG#1{\FIGNUM#1
   \immediate\write\figurewrite{\noexpand\refitem{Fig.#1.}}%
   \begingroup\obeyendofline\rw@start}
\def\figout{\par \penalty-400 \vskip\chapterskip
   \spacecheck\referenceminspace \immediate\closeout\figurewrite
   \figureopenfalse
   \line{\fourteenrm\hfil FIGURE CAPTIONS\hfil}\vskip\headskip
   \input figures.aux
   }
\def\fig{\FIGNUM\? fig.~\?%
\immediate\write\figurewrite{\noexpand\refitem{\?.}}%
\begingroup\obeyendofline\rw@start}
\def\figure{\FIGNUM\? figure~\?
   \immediate\write\figurewrite{\noexpand\refitem{\?.}}%
   \begingroup\obeyendofline\rw@start}
\def\Fig{\FIGNUM\? Fig.~\?%
\immediate\write\figurewrite{\noexpand\refitem{\?.}}%
\begingroup\obeyendofline\rw@start}
\def\Figure{\FIGNUM\? Figure~\?%
\immediate\write\figurewrite{\noexpand\refitem{\?.}}%
\begingroup\obeyendofline\rw@start}
\newcount\tablecount	 \tablecount=0
\newif\iftableopen	 \newwrite\tablewrite
\def\tabch@ck{\chardef\rw@write=\tablewrite \iftableopen\else
   \immediate\openout\tablewrite=tables.aux
   \tableopentrue\fi}
\def\TABNUM#1{\space@ver{}\tabch@ck \firstreflinetrue%
 \global\advance\tablecount by 1 \xdef#1{\the\tablecount}}
\def\TABLE#1{\TABNUM#1
   \immediate\write\tablewrite{\noexpand\refitem{#1.}}%
   \begingroup\obeyendofline\rw@start}
\def\Table{\TABNUM\? Table~\?%
\immediate\write\tablewrite{\noexpand\refitem{\?.}}%
\begingroup\obeyendofline\rw@start}
\def\tabout{\par \penalty-400 \vskip\chapterskip
   \spacecheck\referenceminspace \immediate\closeout\tablewrite
   \tableopenfalse
   \line{\fourteenrm\hfil TABLE CAPTIONS\hfil}\vskip\headskip
   \input tables.aux
   }
%
%
%
\def\masterreset{\global\pagenumber=1 \global\chapternumber=0
   \global\equanumber=0 \global\sectionnumber=0
   \global\referencecount=0 \global\figurecount=0 \global\tablecount=0 }
\def\FRONTPAGE{\ifvoid255\else\vfill\penalty-2000\fi
      \masterreset\global\frontpagetrue
      \global\lastf@@t=0 \global\footsymbolcount=0}

\def\paperstyle{\letterstylefalse\normalspace\papersize}
\def\letterstyle{\letterstyletrue\singlespace\lettersize}
\def\papersize{\hsize=35pc\vsize=50pc\hoffset=1pc\voffset=6pc
		\skip\footins=\bigskipamount}
\def\lettersize{\hsize=6in\vsize=8.5in\hoffset=0.33in\voffset=1in
   \skip\footins=\smallskipamount \multiply\skip\footins by 3 }
\paperstyle   
%
%
\def\MEMO{\letterstyle\FRONTPAGE \letterfrontheadline={\hfil}
    \line{\quad\fourteenrm KEK MEMORANDUM\hfil\twelverm\the\date\quad}
    \medskip \memod@f}

\def\memit@m#1{\smallskip \hangafter=0 \hangindent=1in
      \Textindent{\caps #1}}
\def\memod@f{\xdef\to{\memit@m{To:}}\xdef\from{\memit@m{From:}}%
     \xdef\topic{\memit@m{Topic:}}\xdef\subject{\memit@m{Subject:}}%
     \xdef\rule{\bigskip\hrule height 1pt\bigskip}}
\memod@f
%

\def\nohead{
 \def\letters{\letterstyle \letterfrontheadline={\hfil}}
 \def\letter{\FRONTPAGE\BLANKHEAD\addressee}
}
\nohead
%


%
\def\BLANKHEAD{\hrule height 0pt depth 0pt \vskip 1cm plus .5cm minus .5cm}
\newskip\lettertopfil
\lettertopfil = 0pt plus 1.5in minus 0pt
\newskip\letterbottomfil
\letterbottomfil = 0pt plus 2.3in minus 0pt
\newskip\spskip \setbox0\hbox{\ } \spskip=-1\wd0
\def\addressee#1{\medskip \line{\hskip 0.5\hsize \hbox{\the\date}\hfil}
   \bigskip
   \vskip\lettertopfil
   \ialign to\hsize{\strut ##\hfil\tabskip 0pt plus \hsize \cr #1\crcr}
   \medskip\noindent\hskip\spskip}
\newskip\signatureskip	     \signatureskip=40pt
\def\signed#1{\par \penalty 9000 \bigskip \dt@pfalse
  \everycr={\noalign{\ifdt@p\vskip\signatureskip\global\dt@pfalse\fi}}
  \setbox0=\vbox{\singlespace \halign{\tabskip 0pt \strut ##\hfil\cr
   \noalign{\global\dt@ptrue}#1\crcr}}
  \line{\hskip 0.5\hsize minus 0.5\hsize \box0\hfil} \medskip }

\def\endletter{\ifnum\pagenumber=1 \vskip\letterbottomfil\supereject
\else \vfil\supereject \fi}
\newbox\letterb@x
\def\lettertext{\par\unvcopy\letterb@x\par}
\def\multiletter{\setbox\letterb@x=\vbox\bgroup
      \everypar{\vrule height 1\baselineskip depth 0pt width 0pt }
      \singlespace \topskip=\baselineskip }
\def\letterend{\par\egroup}
%
%
\newskip\frontpageskip
\newtoks\pubtype
\newtoks\Pubnum \newtoks\pubnum
\newtoks\Thnum \newtoks\thnum
\newtoks\s@condpubnum \newtoks\th@rdpubnum
\newif\ifs@cond \s@condfalse
\newif\ifth@rd \th@rdfalse
\newif\ifp@bblock  \p@bblocktrue
\newcount\Year
\def\Yearset{\Year=\year \advance\Year by -1900
 \ifnum\month<4 \advance\Year by -1 \fi}
\def\PH@SR@V{\doubl@true \baselineskip=24.1pt plus 0.2pt minus 0.1pt
	     \parskip= 3pt plus 2pt minus 1pt }
\def\PHYSREV{\paperstyle\PhysRevtrue\PH@SR@V}
\def\titlepage{\Yearset\FRONTPAGE\paperstyle\ifPhysRev\PH@SR@V\fi
   \ifp@bblock\p@bblock\fi}
\def\nopubblock{\p@bblockfalse}
\def\endpage{\vfil\break}
\frontpageskip=1\medskipamount plus .5fil
\pubtype={\tensl Preliminary Version}
\newtoks\publevel
\publevel={Preprint}   
\Pubnum={KEK \the\publevel\ \the\Year--\the\pubnum }
\pubnum={ }
\Thnum={KEK--TH--\the\thnum }
\thnum={ }
\def\secondpubnum#1{\s@condtrue\s@condpubnum={#1}}
\def\thirdpubnum#1{\th@rdtrue\th@rdpubnum={#1}}
\def\p@bblock{\begingroup \tabskip=\hsize minus \hsize
   \baselineskip=1.5\ht\strutbox \topspace-2\baselineskip
   \halign to\hsize{\strut ##\hfil\tabskip=0pt\crcr
   \the\Thnum\cr \the\Pubnum\cr
   \ifs@cond \the\s@condpubnum\cr\fi
   \ifth@rd \the\th@rdpubnum\cr\fi
   \the\Month \cr}\endgroup}
%
\def\title#1{\hrule height0pt depth0pt
   \vskip\frontpageskip \titlestyle{#1} \vskip\headskip }
%
\def\author#1{\vskip\frontpageskip\titlestyle{\twelvecp #1}\nobreak}

\def\address#1{\par\kern 5pt\titlestyle{\twelvepoint\it #1}}
\def\andaddress{\par\kern 5pt \centerline{\sl and} \address}

%
\def\abstract{\vskip\frontpageskip\centerline{\fourteenrm ABSTRACT}
	      \vskip\headskip }

%
%
%
     
     \def\cf{\hbox{\it cf.}}

\def\\{\relax\ifmmode\backslash\else$\backslash$\fi}
\def\globaleqnumbers{\relax\if\equanumber<0\else\global\equanumber=-1\fi}
\def\nextline{\unskip\nobreak\hskip\parfillskip\break}

\def\journal#1&#2(#3){\unskip, \sl #1~\bf #2 \rm (19#3) }

\def\topspace{\hrule height 0pt depth 0pt \vskip}

\let\int=\intop		
\def\prop{\mathrel{{\mathchoice{\pr@p\scriptstyle}{\pr@p\scriptstyle}{
		\pr@p\scriptscriptstyle}{\pr@p\scriptscriptstyle} }}}
\def\pr@p#1{\setbox0=\hbox{$\cal #1 \char'103$}
   \hbox{$\cal #1 \char'117$\kern-.4\wd0\box0}}
\def\lsim{\mathrel{\mathpalette\@versim<}}
\def\gsim{\mathrel{\mathpalette\@versim>}}
\def\@versim#1#2{\lower0.2ex\vbox{\baselineskip\z@skip\lineskip\z@skip
  \lineskiplimit\z@\ialign{$\m@th#1\hfil##\hfil$\crcr#2\crcr\sim\crcr}}}
%
%
%
\let\sec@nt=\sec
\def\sec{\relax\ifmmode\let\n@xt=\sec@nt\else\let\n@xt\section\fi\n@xt}
\def\obsolete#1{\message{Macro \string #1 is obsolete.}}
\def\firstsec#1{\obsolete\firstsec \section{#1}}
\def\firstsubsec#1{\obsolete\firstsubsec \subsection{#1}}
\def\thispage#1{\obsolete\thispage \global\pagenumber=#1\frontpagefalse}
\def\thischapter#1{\obsolete\thischapter \global\chapternumber=#1}
\def\nextequation#1{\obsolete\nextequation \global\equanumber=#1
   \ifnum\the\equanumber>0 \global\advance\equanumber by 1 \fi}
\def\BOXITEM{\afterassigment\B@XITEM\setbox0=}
\def\B@XITEM{\par\hangindent\wd0 \noindent\box0 }
%

%
\catcode`@=12 
\message{ by V.K.}
\everyjob{\input myphyx }
\font\elevenrm=cmr10 scaled\magstephalf
\font\elevenbf=cmbx10 scaled\magstephalf
\font\eleveni=cmmi10 scaled\magstephalf \skewchar\eleveni='177
\font\elevensy=cmsy10 scaled\magstephalf \skewchar\elevensy='60
\font\elevenex=cmex10 scaled\magstephalf
\font\elevensl=cmsl10 scaled\magstephalf
\font\elevenit=cmti10 scaled\magstephalf
\font\eleventt=cmtt10 scaled\magstephalf
\font\elevencp=cmcsc10 scaled\magstephalf
\catcode\lq@=11 %
\def\elevenpoint{\relax
 \textfont0=\elevenrm \scriptfont0=\ninerm \scriptscriptfont0=\sixrm
 \def\rm{\fam0 \elevenrm \f@ntkey=0}\relax
 \textfont1=\eleveni \scriptfont1=\ninei \scriptscriptfont1=\sixi
 \def\oldstyle{\fam1 \eleveni\f@ntkey=1}\relax
 \textfont2=\elevensy \scriptfont2=\ninesy \scriptscriptfont2=\sixsy
 \textfont3=\elevenex \scriptfont3=\elevenex \scriptscriptfont3=\elevenex
 \def\it{\fam\itfam \elevenit \f@ntkey=4 }\textfont\itfam=\elevenit
 \def\sl{\fam\slfam \elevensl \f@ntkey=5 }\textfont\slfam=\elevensl
 \scriptfont\slfam=\ninesl
 \def\bf{\fam\bffam \elevenbf \f@ntkey=6 }\textfont\bffam=\elevenbf
 \scriptfont\bffam=\ninebf \scriptscriptfont\bffam=\sixbf
 \def\tt{\fam\ttfam \eleventt \f@ntkey=7 }\textfont\ttfam=\eleventt
 \h@big=9.311\p@ \h@Big=12.6\p@ \h@bigg=15.88\p@ \h@Bigg=19.17\p@
 \def\caps{\fam\cpfam \elevencp \f@ntkey=8 }\textfont\cpfam=\elevencp
 \setbox\strutbox=\hbox{\vrule height 9pt depth 4pt width\z@}\relax
 \samef@nt}
\newif\ifelev@n \elev@nfalse
\def\Tenpoint{\tenpoint\twelv@false\elev@nfalse\spaces@t}
\def\Elevenpoint{\elevenpoint\twelv@false\elev@ntrue\spaces@t}
\def\Twelvepoint{\twelvepoint\twelv@true\elev@nfalse\spaces@t}
\def\spaces@t{\relax
\iftwelv@ \ifsingl@\subspaces@t3:4;\else\subspaces@t1:1;\fi
\else \ifelev@n \ifsingl@\subspaces@t2:3;\else\subspaces@t9:10;\fi
\else \ifsingl@\subspaces@t3:5;\else\subspaces@t4:5;\fi \fi \fi
\ifdoubl@ \multiply\baselineskip by 5
\divide\baselineskip by 4 \fi}
\catcode\lq @=12 %
\catcode\lq@=11
%
\def\keepspacefactor{\let\@sf=\empty \ifhmode
 \edef\@sf{\spacefactor=\the\spacefactor\relax}\relax\fi}
\newcount\footcount \footcount=0
\def\Footnote{\global\advance\footcount by 1 \footnote{\the\footcount}}
\def\footnote#1{\keepspacefactor\refattach{#1}\vfootnote{#1}}

\def\nonfrenchspacing{\sfcode\lq\.=3000 \sfcode\lq\?=3001 \sfcode\lq\!=3001
 \sfcode\lq\:=2000 \sfcode\lq\;=1500 \sfcode\lq\,=1250 }

\nonfrenchspacing
\newmuskip\refskip
\newmuskip\regularrefskip \regularrefskip=2mu
\newmuskip\specialrefskip \specialrefskip=-2mu
\def\refattach#1{\@sf \ifhmode\ifnum\spacefactor=1250 \refskip=\specialrefskip
 \else\ifnum\spacefactor=3000 \refskip=\specialrefskip
 \else\ifnum\spacefactor=1001 \refskip=\specialrefskip
 \else \refskip=\regularrefskip \fi\fi\fi
 \else \refskip=\regularrefskip \fi
 \ref@ttach{\strut^{\mkern\refskip #1}}}
\def\ref@ttach#1{\ifmmode #1\else\ifhmode\unskip${}#1$\relax\fi\fi{\@sf}}
\def\PLrefmark#1{ [#1]{\@sf}}
\def\NPrefmark#1{\refattach{\scriptstyle [ #1 ] }}
\let\PRrefmark=\refattach
\def\refmark{\keepspacefactor\refm@rk}
%
%
\def\refm@rk#1{\relax\therefm@rk{#1}}
\def\originalrefs{\let\therefm@rk=\NPrefmark}
\def\PRrefs{\let\therefm@rk=\PRrefmark \let\therefitem=\PRrefitem}
\def\PLrefs{\let\therefm@rk=\PLrefmark \let\therefitem=\PLrefitem}
\def\PRrefitem#1{\refitem{#1.}}
\def\PLrefitem#1{\refitem{[#1]}}
\let\therefitem=\PRrefitem
\def\REF#1{\REFNUM#1%
 \immediate\write\referencewrite{%
 \noexpand\therefitem{#1}}%
\begingroup\obeyendofline\rw@start}
\def\ref{\refnum\?%
 \immediate\write\referencewrite{\noexpand\therefitem{\?}}%
\begingroup\obeyendofline\rw@start}
\def\Ref#1{\refnum#1%
 \immediate\write\referencewrite{\noexpand\therefitem{#1}}%
\begingroup\obeyendofline\rw@start}
\def\REFS#1{\REFNUM#1\global\lastrefsbegincount=\referencecount
\immediate\write\referencewrite{\noexpand\therefitem{#1}}%
\begingroup\obeyendofline\rw@start}
\def\refend{\refm@rk{\number\referencecount}}
{\obeyendofline\gdef\rw@start#1{\def\t@st{#1}\ifx\t@st\blankend%
\endgroup {\@sf} \relax \else \ifx\t@st\bl@nkend \endgroup {\@sf} \relax%
\else \rw@begin#1
\backtotext
\fi \fi } }
\refindent=20pt
\def\REFNUM#1{\eatspace\keepspacefactor\refch@ck \firstreflinetrue%
 \global\advance\referencecount by 1 \xdef#1{\the\referencecount}}
\def\eatspace{\ifhmode\unskip\fi}
\def\refnum#1{\keepspacefactor\refch@ck \firstreflinetrue%
 \global\advance\referencecount by 1 \xdef#1{\the\referencecount}\refend}
\def\figout{\par \penalty-400 \vskip\chapterskip
  \spacecheck\referenceminspace \immediate\closeout\figurewrite
  \figureopenfalse
  \line{\fourteenrm
   \hfil FIGURE CAPTION\ifnum\figurecount=1 \else S \fi\hfil}
  \vskip\headskip
  \input figures.aux
  }
\def\figitem#1{\par\indent \hangindent2\parindent \textindent{Fig. #1\ }}
\def\FIGLABEL#1{\ifnum\number#1<10 \def\figlabel{#1.\zerophant}\else%
\def\figlabel{#1.}\fi}
\def\FIG#1{\FIGNUM#1\FIGLABEL#1%
\immediate\write\figurewrite{\noexpand\figitem{\figlabel}}%
\begingroup\obeyendofline\rw@start}
\def\Figname#1{\FIGNUM#1Fig.~#1\FIGLABEL#1%
\immediate\write\figurewrite{\noexpand\figitem{\figlabel}}%
\begingroup\obeyendofline\rw@start}
\def\fig{\FIGNUM\? fig.~\? \FIGLABEL\?
\immediate\write\figurewrite{\noexpand\figitem{\figlabel}}%
\begingroup\obeyendofline\rw@start}
\def\figure{\FIGNUM\? figure~\? \FIGLABEL\?
\immediate\write\figurewrite{\noexpand\figitem{\figlabel}}%
\begingroup\obeyendofline\rw@start}
\def\Fig{\FIGNUM\? Fig.~\? \FIGLABEL\?
\immediate\write\figurewrite{\noexpand\figitem{\figlabel}}%
\begingroup\obeyendofline\rw@start}
\def\Figure{\FIGNUM\? Figure~\? \FIGLABEL\?
\immediate\write\figurewrite{\noexpand\figitem{\figlabel}}%
\begingroup\obeyendofline\rw@start}
\def\FIGNUM#1{\keepspacefactor\figch@ck \firstreflinetrue%
\global\advance\figurecount by 1 \xdef#1{\the\figurecount}}
\newdimen\digitwidth \setbox0=\hbox{\rm0} \digitwidth=\wd0
\def\zerophant{\kern\digitwidth}
\def\TABNUM#1{\keepspacefactor\tabch@ck \firstreflinetrue%
\global\advance\tablecount by 1 \xdef#1{\the\tablecount}}
\def\tableitem#1{\par\indent \hangindent2\parindent \textindent{Table #1\ }}
\def\TABLE#1{\TABNUM#1\FIGLABEL#1%
\immediate\write\tablewrite{\noexpand\tableitem{\figlabel}}%
\begingroup\obeyendofline\rw@start}
\def\Table{\TABNUM\? Table~\?\FIGLABEL\?%
\immediate\write\tablewrite{\noexpand\tableitem{\figlabel}}%
\begingroup\obeyendofline\rw@start}
\def\tabout{\par \penalty-400 \vskip\chapterskip
  \spacecheck\referenceminspace \immediate\closeout\tablewrite \tableopenfalse
  \line{\fourteenrm\hfil TABLE CAPTION\ifnum\tablecount=1 \else S\fi\hfil}
  \vskip\headskip
  \input tables.aux
  }
\catcode\lq @=12
\PRrefs
%

%
\newif\iffinal \finaltrue
\def\showeqname#1{\iffinal\else\hbox to 0pt{\tentt\kern2mm\string#1\hss}\fi}
\def\showEqname#1{\iffinal\else \hskip 0pt plus 1fill
 \hbox to 0pt{\tentt\kern2mm\string#1\hss}\hskip 0pt plus -1fill\fi}
\catcode`\@=11
%
%
\def\eqnamedef#1{\relax \ifnum\equanumber<0
  \xdef#1{{\noexpand\rm(\number-\equanumber)}}\global\advance\equanumber by -1
    \else \global\advance\equanumber by 1
      \xdef#1{{\noexpand\rm(\chapterlabel \number\equanumber)}}\fi}
\def\eqnamenewdef#1#2{\relax \ifnum\equanumber<0
 \xdef#1{{\noexpand\rm(\number-\equanumber#2)}}\global\advance\equanumber by -1
    \else \global\advance\equanumber by 1
      \xdef#1{{\noexpand\rm(\chapterlabel \number\equanumber#2)}}\fi}
\def\eqnameolddef#1#2{\relax \ifnum\equanumber<0
     \global\advance\equanumber by 1
 \xdef#1{{\noexpand\rm(\number-\equanumber#2)}}\global\advance\equanumber by -1
    \else \xdef#1{{\noexpand\rm(\chapterlabel \number\equanumber#2)}}\fi}
\def\eqname#1{\eqnamedef{#1}#1}
\def\eqnamenew#1#2{\eqnamenewdef{#1}{#2}#1}
\def\eqnameold#1#2{\eqnameolddef{#1}{#2}#1}
\def\eq{\eqname\lasteq}
\def\eqa{\eqnamenew\lasteq a}
\def\eqb{\eqnameold\lasteq b}
\def\eqc{\eqnameold\lasteq c}
\def\eqd{\eqnameold\lasteq d}
\def\eqnew#1{\eqnamenew\lasteq{#1}}
\def\eqold#1{\eqnameold\lasteq{#1}}
%

%

%
\def\eq@@{\ifinner\let\eqn@=\relax\else\let\eqn@=\eqno\fi\eqn@}
\def\Eq{\eq@@\eq}
\def\Eqnew#1{\eq@@\eqnew{#1}}
\def\Eqold#1{\eq@@\eqold{#1}}
\def\Eqa{\eq@@\eqa}
\def\Eqb{\eq@@\eqb}
\def\Eqc{\eq@@\eqc}
\def\Eqd{\eq@@\eqd}
\def\Eqn#1{\eq@@\eqname{#1}\showeqname{#1}}
\def\Eqnnew#1#2{\eq@@\eqnamenew{#2}{#1}\showeqname{#1}}
\def\Eqnold#1#2{\eq@@\eqnameold{#2}{#1}\showeqname{#1}}
\def\Eqna#1{\eq@@\eqnamenew{#1}a\showeqname{#1}}
\def\Eqnb#1{\eq@@\eqnameold{#1}b\showeqname{#1}}
\def\Eqnc#1{\eq@@\eqnameold{#1}c\showeqname{#1}}
\def\Eqnd#1{\eq@@\eqnameold{#1}d\showeqname{#1}}

\catcode`\@=12
%
\paperheadline={\hfil\iffinal\else\tenrm\the\date\fi}
%
\message{ Modified by K. Hikasa: Jan. 9, 1992 Version (PLrefmark modified)}
%
%
%
%
%
%
%
%
%
%
%
%
%
\def\abstract{\vskip\frontpageskip\centerline
   {\fourteenbf  Abstract} \vskip\headskip }
\def\appendix#1#2{\par\penalty-300\vskip\chapterskip
   \spacecheck\chapterminspace \chapterreset \xdef\chapterlabel{#1}
   \titlestyle{Appendix #2} \nobreak\vskip\headskip \penalty 30000
   \wlog{\string\Appendix\ \chapterlabel} }
\def\Appendix#1{\appendix{#1.}{#1}}
\def\ack{\par\penalty-100\medskip \spacecheck\sectionminspace
   \line{\fourteenbf \hfil  Acknowledgments \hfil}
   \nobreak\vskip\headskip }

\def\p@bblock{\begingroup \tabskip=\hsize minus \hsize
   \baselineskip=1.5\ht\strutbox \topspace-2\baselineskip
   \halign to\hsize{\strut ##\hfil\tabskip=0pt\crcr
   \the\Thnum\cr \the\Pubnum\cr
   \ifs@cond \the\s@condpubnum\cr\fi
   \ifth@rd \the\th@rdpubnum\cr\fi
   \the\Month 1992 \cr}\endgroup}
\overfullrule=0pt
\Thnum={KEK--TH--310}
\Pubnum={KEK Preprint 91--189}
\Month={December 1991}
\titlepage
\title{{\bf Screening Currents Ward Identity          \break \vskip-5mm
                           and                        \break \vskip-5mm
 Integral Formulas for the WZNW Correlation Functions}}
\footnote\dag{Talk given at,
``Workshop on Development in String Theory and New Field Theories'',
Kyoto Sept. 1991.}
\author{ Hidetoshi Awata }
\footnote\ddag{E-mail address : awata@jpnkekvm.bitnet}
\vskip 1cm
\address{ National Laboratory for High Energy Physics {\rm (}KEK{\rm )}
         \nextline Tsukuba, Ibaraki 305, Japan
         \nextline and
         \nextline Dept. of Physics Hokkaido University,
         \nextline Sapporo 060, Japan}
{\bf \abstract}
 We derive, based on the Wakimoto realization,
 the integral formulas for the WZNW correlation functions.
The role of the {\it ``screening currents Ward identity''}
is demonstrated with explicit examples.
We also give a more simple proof of a previous result.
\endpage

\REF\kz{V. Knizhnik and A. Zamolodchikov,
           Nucl. Phys. {\bf B247} (1984) 83 }

\REF\tk{A. Tsuchiya and Y. Kanie,
         Lett. Math. Phys. {\bf 13} (1987) 303,
         Advanced Studies in Pure Math. {\bf 16} (1988) 297}

\REF\gw{D. Gepner and E. Witten, Nucl. Phys. {\bf B278} (1986) 493}

\REF\wa{M. Wakimoto, Comm. Math. Phys. {\bf 104} (1986) 605}

\REF\fz{V.A. Fateev and A.B. Zamolodchikov,
          Sov. J. Nucl. Phys. {\bf 43} (4) (1986) 657}

\REF\bo{M. Bershadsky and H. Ooguri,
          Comm. Math. Phys. {\bf 126} (1989) 49}

\REF\bef{D. Bernard and G. Felder,
        Comm. Math. Phys. {\bf 127} (1990) 145}

\REF\gmmos{A. Gerasimov, A. Marshakov, A. Morozov, M. Olshanetskii
           and S. Shatashvili,
		   Int. J. Mod. Phys. {\bf A5} (1990) 2495  }

\REF\ffr{B.L. Feigin  and E.V. Frenkel,
		 Comm. Math. Phys. {\bf 128} (1990) 161}

\REF\bmp{P. Bouwknegt, J. McCarthy and K. Pilch,
           Phys. Lett. {\bf B234} (1990) 297,
           Comm. Math. Phys. {\bf 131} (1990) 125}

\REF\kosik{M. Kuwahara, N. Ohta and H. Suzuki,
		   Phys. Lett. {\bf B235} (1990) 57,
           Nucl. Phys. {\bf B340} (1990) 448; \hfill\break

		   K. Ito and Y. Kazama,
           Mod. Phys. Lett. {\bf A5} (1990) 215}

\REF\iko{  K. Ito and S. Komata,
           Mod. Phys. Lett. {\bf A6} (1991) 581}

\REF\ito{K. Ito, YITP preprint YITP/K-885 (1990)}

\REF\dot{Vl.S. Dotsenko, Nucl. Phys. {\bf B338} (1990) 747 }

\REF\ku{G. Kuroki, Tohoku Univ. preprint December (1990)}

\REF\ao{K. Aomoto, J. Math. Soc. Japan {\bf 39} (2) (1987) 191}

\REF\cf{P. Christe and R. Flume, Nucl. Phys.  {\bf B282} (1987) 466}

\REF\djmm{E. Date, M. Jimbo, A. Matsuo and T. Miwa,
           Int. J. Mod. Phys. {\bf B4} (1990) 1049}

\REF\ma{A. Matsuo, Comm. Math. Phys. {\bf 134 } (1990) 65 }

\REF\sv{V.V. Schechtman and A.N. Varchenko,
        {\it Integral representations of n-point conformal correlators
		in the WZW model},
		preprint MPI/89-51 (1989);\hfill\break

		see also Lett. Math. Phys. {\bf 20} (1990) 279}

\REF\aty{H. Awata, A. Tsuchiya and Y. Yamada,
          Nucl. Phys. {\bf B365 } (1991) 680 }

\REF\df{ Vl.S. Dotsenko and V.A. Fateev,
          Nucl. Phys.  {\bf B240} [FS12] (1984) 312,
		               {\bf B251} [FS13] (1985) 691}

\REF\fe{G. Felder, Nucl. Phys.  {\bf B317} (1989) 215,
        erratum in {\bf B324} (1989) 548}

\REF\qg{V.G. Drinf'eld, Int. Cong. of Math. (1986) p.798;\hfill\break

	 M. Jimbo, Lett. Math. Phys. {\bf 10} (1985) 63;\hfill\break

	 L. Faddeev, N. Yu. Reshetikhin, and L. A. Takhatajian,
	 LOMI preprint E-14-87, Algebre i Analiz {\bf 1} (1989) 87}

\REF\sva{V.V. Schechtman and A.N. Varchenko, Moscow preprint
     January and August (1990)}

\REF\ay{H. Awata and Y. Yamada,
       {\it Fusion rules for
	   the fractional level $\widehat{sl(2)}$ algebra},
		KEK Preprint 91-209, KEK-TH-316  }

{\bf \chapter{Introduction}}

 In the Wess-Zumino-Novikov-Witten (WZNW) models [\kz $-$\gw],
the correlation function can be characterized as the solution of
the Knizhnik-Zamolodchikov (KZ) equation.
The integral representation for this correlation function
has been studied in the last few years [\fz, \cf $-$\sv].

 It is known that this representation can be obtained from
the free field realization of the Kac-Moody algebras
(the Wakimoto realization).
\footnote\dag{In [\djmm $-$\sv],
they constructed this integral representation by another approach
based on the generalized theory of hypergeometric type equations [\ao].
}
 This realization has been discussed by several authors [\wa $-$\bo],
and its general properties has been clarified [\bef $-$\bmp].
In this method, the correlation function is represented by some
integral of the free fields correlation.
But, for arbitrary algebras and representations,
it is difficult to evaluate the free fields
correlation except in some particular cases [\gmmos, \dot].
 This is entirely due to the complicated structure of the realization
[\kosik $-$\ito].

 In a previous paper [\aty], we solved this problem and derived the
integral representation for the WZNW models corresponding to arbitrary
simple Lie algebras.
By using the {\it ``screening currents Ward identity''},
we obtained a exact result without treating the complicated explicit
form of the Wakimoto realization.

This paper is a more complete version of [\aty],
and contains some explicit examples in the case of $\widehat{sl(3)}$
and refined proofs.

 This paper is arranged as follows.
In section 2, we refer to  the free field realization of
finite dimensional simple Lie algebras.
In section 3, the Wakimoto realization of the affine Kac-Moody algebras is
introduced.
Next, in section 4, we study some examples of the integral formulas
for the WZNW correlation functions.
We then present the systematic derivation of these integral formulas
in section 5.
This section contains the main results of the paper.
Some of the proofs are given in the Appendices.

{\bf \chapter{The Free Field Realization of Simple Lie Algebras}}

 We start with recapitulating the results of
the free field realization
( differential realization ) of simple Lie algebras.
\vskip 2mm
\noindent{\bf \S$\,$2.1.}~
 A finite dimensional simple Lie algebra ${\bf g}$
is defined by the following commutation relations
for the Chevalley generators,
$e_{\alpha_i}$,$f_{\alpha_i}$ and $h_i$ $(i=1, \cdots,l=$rank${\bf g})$
$$\eqalign{
 [ h_i, h_j ] &= 0,                             \hskip 1.8cm
 [ e_{\alpha_i}, f_{\alpha_j} ] = \delta_{ij} h_j, \cr
 [ h_i, e_{\alpha_j} ] &=  a_{ij} e_{\alpha_j}, \hskip 1.1cm
 [ h_i, f_{\alpha_j} ] = -a_{ij} f_{\alpha_j}. \cr
}\eqno\eq$$
where $\alpha_i$ is the simple root, and $a_{ij}$ is the Cartan matrix. The
Cartan matrix is realized as $a_{ij}=(\nu_i \cdot \alpha_j)$,
where $\nu_i={2 \over \alpha_i^2}\alpha_i$ is the coroot of $\alpha_i$, and (
$\cdot$ ) is the symmetric bilinear form.
\footnote\dag{For the quantities in Cartan space {\bf h},
for example $h$, $H$, $\lambda$, $\phi$,
their components are defined by
$\lambda_i=(\nu_i \cdot \lambda)$ and
$\lambda=\sum_i \lambda^i \nu_i$ etc.}

 The Verma module $V_{\lambda}$ is generated by
the highest weight vector $\vert \lambda \rangle $ which satisfies
$e_{\alpha} \vert \lambda \rangle = 0$,
$h_i        \vert \lambda \rangle = \lambda_i \vert \lambda \rangle $.
The Dual module $V_\lambda^*$ is generated by
$\langle \lambda \vert $ such that
$\langle \lambda \vert f_{\alpha} = 0$,
$\langle \lambda \vert h_i        = \lambda_i \langle \lambda \vert $.
 The bilinear form
$V_\lambda^* \otimes V_\lambda \rightarrow {\bf C}$
defined by $\langle \lambda \vert \lambda \rangle  =1$
is called the Shapovalov form.
\footnote{\ddag}{
 Strictly speaking the Shapovalov form is a bilinear form on
$V_\lambda \otimes V_\lambda$. }

\vskip 2mm
\noindent{\bf \S$\,$2.2.}~
 The algebra ${\bf g}$ is realized on the polynomial ring
${\bf C} [x^\alpha ]$, with positive root $\alpha \in \Delta_+$,
as (twisted first order) differential operators.
 The differential operators
$J({\partial \over \partial x},x,\lambda )$'s
corresponding to the generators $J$'s are defined  by the following
``right action''
\footnote{\star} {For a positive generator $e_\alpha $,
$E_\alpha ({\partial \over \partial x},x )$
can be defined as
$ E_\alpha ({\partial \over \partial x},x ) \, Z
= Z \, e_\alpha  $  .}
$$ J({\partial \over \partial x},x,\lambda ) \langle \lambda \vert  \, Z
=\langle \lambda \vert  \, Z \,  J  ,
\eqno\eq  $$
where $ Z \equiv \exp ( \sum_\alpha x^\alpha e_\alpha )  $.
They are given by
$$J({\partial \over \partial x},x,\lambda )
=\sum_{\alpha > 0 }
  V^\alpha(x) {\partial \over \partial x^\alpha}
+\sum_{i=1}^l
  W^i(x)       \lambda_i ,
\eqno\eq$$
for some polynomials
$V^\alpha(x)$ and  $W^i(x) \in {\bf C} [x] $.

 There is another type of differential operators $S_{\alpha}$
induced by ``left action'' of $e_{\alpha}$
that plays an important role in this paper.
They are defined by
$$ S_{\alpha}({\partial \over \partial x},x )  \, Z
       = - e_{\alpha} \, Z ,
\eqno\eq  $$
and with some polynomials $S_\alpha^\beta(x)$ as
$$S_\alpha({\partial \over \partial x},x )
=\sum_{\beta  > 0 }
  S_\alpha^\beta(x) {\partial \over \partial x^\beta} .
\eqno\eq$$
{}From the associativity of
$\langle \lambda \vert  \, e_\alpha \, Z \,  J $ ,
we have
\footnote{\dag}{The formula in [\aty] has a wrong sign.}
$$\eqalign{[E_\alpha,S_\beta]&=0,\cr
[H_i     ,S_\alpha]&=(\nu_i \cdot \alpha)S_\alpha,\cr
[F_{\alpha_i},S_{\alpha_j}]
&=(\nu_i \cdot \alpha_j) \, x^{\alpha_i}S_{\alpha_j}
+\lambda_i \delta_{ij} .}\eqno\eq$$
Useful formulas and the proof of (2.6) are given in Appendix A.

\noindent {\bf EXAMPLE.}~  The $sl(3)$ algebra, whose Cartan matrix is
$a_{11}=a_{22}=2$ and $a_{12}=a_{21}=-1$,
is realized on the polynomial ring
${\bf C} [x^{\alpha}] \; \alpha =1,2,3 $,
where $x^3$ is associated with $e_3=[e_1,e_2]$.
 For example the differential operators $E_1$, $F_1$ and $H_1$
corresponding to the Chevalley generators $e_1 $, $f_1 $ and $h_1 $
are
$$\eqalign{ E_1 &= {\partial \over \partial x^1}
       -{1 \over 2} x^2 {\partial \over \partial x^3}, \cr
 H_1 &= -2 x^1 {\partial \over \partial x^1}
         + x^2 {\partial \over \partial x^2}
		 - x^3 {\partial \over \partial x^3}
		 + \lambda_1, \cr
 F_1 &= - x^1 x^1 {\partial \over \partial x^1}
        + ( {1 \over 2} x^1 x^2 - x^3 ) {\partial \over \partial x^2}
		-   {1 \over 2} x^1 ( {1 \over 2} x^1 x^2 + x^3 )
		    {\partial \over \partial x^3}
		+ \lambda_1 x^1, \cr}\eqno\eq$$
and $S_1$ is
$$S_1 = - {\partial \over \partial x^1}
       -{1 \over 2} x^2 {\partial \over \partial x^3} .\eqno\eq$$
 $E_2$, $F_2$, $H_2$ and $S_2$ are given by replacing
$x^1 $ with $x^2$, $x^3$ with $-x^3 $
and $\lambda_1$ with $\lambda_2$.

\vskip 2mm
\noindent{\bf \S$\,$2.3.}~
The state $\vert 0\rangle  \equiv 1 \in {\bf C} [x^\alpha]$
is the highest weight vector such that
$${\partial \over \partial x^\alpha} \vert 0\rangle  =0, \hskip 1cm
E_\alpha \vert 0\rangle  =0, \hskip 1cm
H_\mu \vert 0\rangle  = \lambda_\mu \vert 0\rangle .\eqno\eq$$
For an ordered set $I=\{\alpha_1,\cdots,\alpha_n\}$,
\footnote\dag{We sometimes use the over-simplified notation
$I=\{\alpha_1,\cdots,\alpha_n\}$ to denote
an ordered set $I$ of simple roots ${\alpha_i}$,
although they should be understood as
$\{\alpha_{i_1},\cdots,\alpha_{i_n}\}$.  }
the vectors $P_\lambda^I \vert 0 \rangle
=\prod_{\alpha_i\in I} F_{\alpha_i} \vert 0\rangle
\in {\bf C}[x^\alpha]$
form the basis of the descendants of $\vert 0 \rangle$.
\footnote\ddag{
 In what follows, the operator factors in the product symbol $\prod$
are assumed to be ordered in the sense that
$\prod_{i\in \{ 2,1,1 \}}{\bf O}_i={\bf O}_2{\bf O}_1{\bf O}_1
\hskip 0.5cm (\neq {\bf O}_1{\bf O}_1{\bf O}_2) . $}
{}From (2.2),
this basis can be written as the following expectation values
$$P_\lambda^I
= \langle \lambda \vert Z
\prod_{i=1}^n f_{\alpha_i} \vert \lambda \rangle.\eqno\eq$$

\noindent{\bf  EXAMPLE.}~
The first few examples for $sl(3)$ are
$$\eqalign{P_{\lambda}^{\{1\}}    &= \lambda_1 x^1,\cr
P_{\lambda}^{\{1,1\}}  &= \lambda_1 (\lambda_1 - 1) x^1 x^1,\cr
P_{\lambda}^{\{2,1\}}
&= \lambda_1 ( ( \lambda_2 + { 1 \over 2}) x^1 x^2 + x^3 ),\cr
P_{\lambda}^{\{1,1,1\}}
&= \lambda_1 (\lambda_1 - 1) (\lambda_1 - 2) x^1 x^1 x^1,\cr
P_{\lambda}^{\{2,1,1\}}
&= \lambda_1 (\lambda_1 - 1)( ( \lambda_2 + 1 ) x^1 x^2 + 2x^3) x^1,\cr
P_{\lambda}^{\{1,2,1\}}
&= \lambda_1 (
(( \lambda_1 - { 1 \over 2}) ( \lambda_2 + { 1 \over 2}) - {1 \over 4})
 x^1 x^2 + ( \lambda_1 - \lambda_2 - 1 ) x^3 ) x^1,\cr
P_{\lambda}^{\{1,1,2\}}
&= \lambda_1 \lambda_2 ( \lambda_1 x^1 x^2 - 2 x^3 ) x^1.}\eqno\eq$$

 Since we have the Serre relations,
this basis is not linearly independent.
For instance,
$$   P_{\lambda}^{\{1,1,2\}} -2 P_{\lambda}^{\{1,2,1\}}
+  P_{\lambda}^{\{2,1,1\}} =0 .\eqno\eq$$

{\bf \chapter
{The Wakimoto Realization }}

 Next we turn to the Wakimoto realization of
the affine Kac-Moody algebras.
\vskip 2mm
\noindent{\bf \S$\,$3.1.}~
 The affine Kac-Moody algebra $\hat {\bf g}$ associated with the Lie algebra
${\bf g}$ is defined by the operator product expansions (OPE);
$$\eqalign{
H_i(z) H_j(w)& ={k \over (z-w)^2}(\nu_i \cdot \nu_j) + \cdots, \cr
H_i(z) E_{\alpha_j}(w)& = {1 \over z-w} a_{ij}E_{\alpha_j}(w) + \cdots, \cr
H_i(z) F_{\alpha_j}(w)& = {-1 \over z-w} a_{ij}F_{\alpha_j}(w) + \cdots,\cr
E_{\alpha_i}(z)F_{\alpha_j}(w)& ={k \over (z-w)^2}{2 \over \alpha_i^2}
 \delta_{ij} + {1 \over z-w}\delta_{ij}H_j(w) + \cdots,}\eqno\eq$$
 where $k$ is the level.
 The energy-momentum tensor $T_{Sug}(z)$ is given by the Sugawara construction;
$$T_{Sug}(z)=(2\kappa)^{-1}
     : \sum_{i=1}^l H_i(z) H^i(z) +
       \sum_{\alpha >0} {{\alpha}^2 \over 2}
	   \big( E_{\alpha}(z)F_{\alpha}(z)+F_{\alpha}(z)E_{\alpha}(z) \big)
:,\eqno\eq$$
where $\kappa=k+h$ and $h$ is the dual Coxeter number of ${\bf g}$.

 For each positive root $\alpha \in \Delta_+$, we introduce bosons
$\beta_{\alpha}(z)$ and $\gamma^{\alpha}(z)$, with conformal weights 1 and 0,
satisfying the canonical OPE
$$\beta_{\alpha}(z) \gamma^{\beta}(w) ={1 \over {z-w}} \delta_{\alpha}^{\beta}
+\cdots, \eqno \eq$$
 We also introduce free bosons  $\phi_i(z)$'s for $i=1, \cdots ,l$, with the
OPE
\footnote{\dag} {
Our notation for $\phi(z)$ is different from the ordinary one.
To get the ordinary $\phi(z)$, we must substitute $\phi(z) \rightarrow i \,
\kappa^{-1/2} \phi(z)$.}
$$\phi_i(z) \phi_j(w) =\kappa^{-1}  (\nu_i \cdot \nu_j)log(z-w)+\cdots,
\eqno \eq$$

 The Kac-Moody algebra $(3.1)$ can be realized
by the free fields $(3.3)$ and $(3.4)$.

\noindent{\bf PROPOSITION.}
 {\it$\,$ The free field realization of the Kac-Moody currents $J(z)$
is given by $[\wa -\ku]$
$$J(\beta,\gamma,\kappa \partial \phi)(z)=:\sum_{\alpha  > 0 }
  V^\alpha(\gamma) \beta_\alpha
+\kappa \sum_{i=1}^l   W^i(\gamma) \partial \phi_i
+\sum_{\alpha  > 0 } A_\alpha (\gamma) \partial \gamma^\alpha :(z)
\eqno\eq$$
where
$V^\alpha \big( \gamma(z) \big) $ and  $W^i \big( \gamma(z) \big) $
$\in {\bf C}[\gamma^\alpha(z)] $
are the operators corresponding to the polynomials
$V^\alpha(x) $ and  $W^i(x)$ $\in {\bf C}[x^\alpha]$ of $(2.3)$ ;
and $A_\alpha \big( \gamma(z) \big) $
are the anomalous parts, due to the normal ordering,
which vanish for $E_\alpha(z)$ and $H_i(z)$.
For $F_{\alpha_i}(z)$, this is given by
$$A_\alpha(\gamma(z)) = \delta_{\alpha_i \alpha}
({2 \over {\alpha_i}^2}k + {h-{\alpha_i}^2 \over {\alpha_i}^2}).
\eqno\eq$$
}

The energy-momentum tensor $T_{Sug}(z) $ is also realized as
$$T_{Sug}(z)
= \sum_{\alpha >0} : \partial \gamma^{\alpha}\beta_{\alpha}:(z)
+ \sum_{i=1}^l
:{\kappa \over 2} \partial \phi_i \partial \phi^i
- \rho_i \partial^2 \phi^i:(z),\eqno\eq$$
where $\rho={1 \over 2} \sum_{\alpha >0} \alpha$ is the half sum of positive
roots.

\noindent{\bf \S$\,$3.2.}~
 There exists one more important ingredient in the Wakimoto realization, that
is the screening current.
Define the screening current $s_i(z)$,
corresponding to the simple root $\alpha_i$
$$\eqalign{s_i(z)&=S_{\alpha_i}(z) e^{-{\alpha_i} \cdot \phi(z)},\cr
\noalign{\vskip 3mm}
 S_{\alpha}(z) &= \sum_{\beta  > 0 } : S_{\alpha}^{\beta}
 \big( \gamma(z) \big) \beta_{\beta}(z) :,}\eqno\eq$$
 where the polynomial $S_{\alpha}^{\beta}(x) \in {\bf C}[x^\alpha]$
is the same as in (2.5).

\endpage

Then we get

\noindent{\bf PROPOSITION.}
{\it $\,$  The screening current $s_i(z)$ satisfies }
$$\eqalign{ E_{\alpha}(z)s_j(w) &= 0 +\cdots, \cr
 H_i(z)s_j(w) &= 0 +\cdots, \cr
 F_{\alpha_i}(z)s_j(w) &= -\kappa \delta_{ij}{2  \over {\alpha_i}^2}
 {\partial \over \partial w}
 \big( {1 \over z-w} e^{-\alpha_j \cdot \phi(w)} \big) +\cdots,\cr
\noalign{\vskip 3mm}
 T_{Sug}(z)s_j(w) &=  {\partial \over \partial w}
 \big( {1 \over z-w} s_j(w) \big) +\cdots.}\eqno\eq$$

 Note that the screening charge $\int dt e^{-\alpha \cdot
\phi(t)}S_{\alpha}(t)$ is well defined only for the simple roots $\alpha_i$'s,
though $S_{\alpha}(t)$ is well-defined for all the positive roots.

\vskip 2mm
\noindent{\bf \S$\,$3.3.}~
 Now we give a natural highest weight representation.
For arbitrary weight vector $\lambda$, the vertex operator  $e^{\lambda \cdot
\phi(z)}$ satisfies the highest weight condition;
$$\eqalign{ E_{\alpha} (z) e^{\lambda \cdot \phi(w)} &= 0+\cdots,  \cr
 H_i(z)e^{\lambda \cdot \phi(w)}
     &= { \lambda_i \over z-w } e^{\lambda \cdot \phi(w)}+\cdots
.}\eqno\eq$$
It has the conformal weight $\lambda \cdot (\lambda +2\rho)/2\kappa$
with respect to the $T_{sug}(z)$.

 For an ``ordered'' set of simple roots, $I=\{\alpha_1,\cdots,\alpha_n\}$,
the fields
$$e^{\lambda \cdot \phi(z)}P_{\lambda}^I(z) \equiv
\int\prod_{k=1}^n {dt_k \over 2\pi i} F_{\alpha_k}(t_k)
e^{\lambda \cdot \phi(z)}\eqno\eq$$
form the basis of the descendants of the highest weight vector
$e^{\lambda \cdot \phi(z)}$.
 Since the $\partial \gamma (z)$ terms (3.5) give no singularity,
they do not contribute to this integral.
So the operator $ P_{\lambda}^I(z) \in {\bf C}[\gamma^\alpha (z)]$
can be constructed from a classical polynomial
$P_\lambda^I \in {\bf C}[x^{\alpha}]$ (2.11) by replacing
$x^{\alpha}$ with $\gamma^{\alpha}(z)$.

Hence the vectors $e^{\lambda \cdot \phi(z)}P_{\lambda}(z)$, with arbitrary
polynomial $P_{\lambda}(z) \in {\bf C}[\gamma^\alpha (z)]$,
form the highest weight representation of the Wakimoto realization.
\footnote\dag{Here we only consider the $L_0$ ground states.}

{\bf \chapter{ Some Examples of Integral Formulas }}

 We study some examples of the integral formulas
for the WZNW correlation functions, based on the Wakimoto realization.
Here we use the $\beta$$\gamma$ OPE relation
and the explicit form for $S_{\alpha}(t)$ and $P(z)$.

\vskip 2mm
\noindent{\bf \S$\,$4.1.}~
 A correlation function of the chiral primary fields
is represented by the correlation of the screening charges
as well as the vertex operators in the form
$$\int\prod_{i=1}^{m}dt_i\langle \prod_{i=1}^{m}e^{-\alpha_i \cdot
\phi(t_i)}S_i(t_i)
        \prod_{a=1}^{n}e^{\lambda_a \cdot \phi(z_a)}P_a(z_a)\rangle .
\eqno \eq$$
 Here $\alpha_1,...,\alpha_m$ are simple roots that fill the gap between the
highest weights of the incoming states $\lambda_1,...,\lambda_n$ and the
outgoing state $\lambda_{\infty}$, i.e.
$$ \lambda_{\infty} = \sum_{a=1}^n \lambda_a -\sum_{i=1}^m \alpha_i.
\eqno \eq$$
 In the set $\{\alpha_1,...,\alpha_m\}$ or $\{\lambda_1,...,\lambda_n\}$ some
$\alpha$'s or $\lambda$'s may be repeated.

 Calculation of the $\phi$ field correlation is given by
the difference products
$$\eqalign{Q& \equiv \langle \prod_{i=1}^{m}e^{-\alpha_i \cdot \phi(t_i)}
      \prod_{a=1}^{n}e^{\lambda_a \cdot \phi(z_a)}\rangle  \cr
  &=\prod_{i < j}^{m}(t_i-t_j)^{{\alpha_i \cdot \alpha_j} \over \kappa}
  \prod_{i=1}^{m}\prod_{a=1}^{n}(t_i-z_a)^{-{\alpha_i \cdot \lambda_a} \over
\kappa}
  \prod_{a < b}^{n}(z_a-z_b)^{{\lambda_a \cdot \lambda_b} \over \kappa}.}\eqno
\eq$$


\noindent{\bf \S$\,$4.2.}~
 Let us next calculate the $\beta\gamma$ correlation
$$\omega \equiv \langle \prod_{i=1}^{m}S_{\alpha_i}(t_i)
 \prod_{a=1}^{n}P_a(z_a)\rangle .\eqno \eq$$
 Since $ S_\alpha(t) $ ( $ P_a (z)$ ) have conformal weight 1 ( 0 ),
this form $\omega$ is a single valued meromorphic 1-form ( function )
on the sphere with respect to the $t$'s ( $z$'s ).

 From the transformation and the singularity property (3.3),
we obtain the following relation
$$\eqalign{\langle
\beta_\alpha(&z)  \prod_{i=1}^{n} \beta_{\alpha_i}(z_i)
                  \prod_{j=1}^{m} \gamma^{\beta_j}(w_j)
				  \rangle  \cr
\noalign{\vskip 2mm}
&= \sum_{k=1}^m {\delta_{\alpha}^{\beta_k} \over {z-w_k}}
\langle           \prod_{i=1}^{n} \beta_{\alpha_i}(z_i)
                  \prod_{j \neq k}^m \gamma^{\beta_j}(w_j)
                  \rangle.}\eqno \eq$$

 This is because both sides,
which are global 1-form on the sphere with respect to $z$,
have the same poles and residues.
 The charge at infinity
which compensates for the $ \beta  \gamma $ current anomaly
is included implicitly.
 From this relation we can  in principle obtain $\omega$.

\noindent{\bf EXAMPLE.}~
Let us consider some examples of $\widehat{sl(3)}$.
Recall that
$ S_1(z) = : - \beta_1 - {1 \over 2} \gamma^2 \beta_3 : (z)$ ,
$ S_2(z) = : - \beta_2 + {1 \over 2} \gamma^1 \beta_3 : (z)$
, and using (2.11), we have
$$\eqalign{
\langle S_1(t) P_{\lambda}^{\{1\}}(z) \rangle
&= \langle -( \beta_1 + {1 \over 2} \gamma^2 \beta_3 )(t)
              \lambda_1 \gamma^1 (z)   \rangle \cr
&= - \lambda_1 \langle \beta_1(t) \gamma^1(z) \rangle \cr
&= - { \lambda_1 \over t - z }, }\eqno\eq$$
$$\eqalign{
\langle S_1(t_1)&S_2(t_2) P_{\lambda}^{\{1,2\}}(z) \rangle \cr
&= \langle ( \beta_1 + {1 \over 2} \gamma^2 \beta_3 )(t_1)
           ( \beta_2 - {1 \over 2} \gamma^1 \beta_3 )(t_2)
 \lambda_2 ( ( \lambda_1 + {1 \over 2} ) \gamma^1 \gamma^2
                                       - \gamma^3 )(z) \rangle \cr
&=   \lambda_2 ( \lambda_1 + {1 \over 2} )
     \langle \beta_1(t_1) \beta_2(t_2) \gamma^1 \gamma^2 (z) \rangle
 + { \lambda_2 \over 2 }
     \langle \beta_1 (t_1) \gamma^1 \beta_3 (t_2) \gamma^3(z)\rangle\cr
&\; \hskip20mm - { \lambda_2 \over 2 }
     \langle \gamma^2 \beta_3 (t_1) \beta_2 (t_2) \gamma^3(z)\rangle\cr
&=   {  \lambda_2 ( \lambda_1 + {1 \over 2} ) \over
       ( t_1 - z )( t_2 - z ) }
   + { \lambda_2 / 2 \over ( t_1 - t_2 )( t_2 - z ) }
   - { \lambda_2 / 2 \over ( t_2 - t_1 )( t_1 - z ) } \cr
&=\big( { -1 \over t_1 - t_2 }
       -{ \lambda_1  \over t_1 - z } \big)
  \big(-{ \lambda_2  \over t_2 - z } \big) ,}\eqno \eq$$
$$\eqalign{
\langle S_1(t_1)& S_1(t_2) P_{\lambda}^{\{1,1\}}(z) \rangle \cr
&= \langle ( \beta_1 + {1 \over 2} \gamma^2 \beta_3 )(t_1)
           ( \beta_1 + {1 \over 2} \gamma^2 \beta_3 )(t_2)
             \lambda_1 ( \lambda_1 - 1 ) \gamma^1 \gamma^1 (z)
		   \rangle \cr
&= \lambda_1 ( \lambda_1 - 1 )
   \langle \beta_1(t_1) \beta_1(t_2) \gamma^1 \gamma^1(z) \rangle \cr
&= { 2 \lambda_1 ( \lambda_1 - 1 ) \over ( t_1 - z ) ( t_2 - z ) }, \cr
&=\big( {         2 \over t_1 - t_2 }
       -{ \lambda_1 \over t_1 - z } \big)
  \big(-{ \lambda_1 \over t_2 - z } \big)
 +\big( {         2 \over t_2 - t_1 }
       -{ \lambda_1 \over t_2 - z } \big)
  \big(-{ \lambda_1 \over t_1 - z } \big) ,}\eqno\eq$$
and
$$\eqalign{\langle S_1(t_1) S_1(t_2)
&P_{\lambda^1}^{\{1\}}(z_1) P_{\lambda^2}^{\{1\}}(z_2) \rangle \cr
&= \langle ( \beta_1 + {1 \over 2} \gamma_2 \beta_3 )(t_1)
           ( \beta_1 + {1 \over 2} \gamma_2 \beta_3 )(t_2)
\,      \lambda^1_1 \gamma_1 (z_1)
		     \lambda^2_1 \gamma_1 (z_2)  \rangle \cr
&= \lambda^1_1 \lambda^2_1
   \langle \beta_1(t_1) \beta_1(t_2)
           \gamma_1(z_1) \gamma_1(z_2) \rangle \cr
&=\big( { \lambda^1_1  \over t_1 - z_1 } \big)
  \big( { \lambda^2_1  \over t_2 - z_2 } \big)
 +\big( { \lambda^1_1  \over t_2 - z_1 } \big)
  \big( { \lambda^2_1  \over t_1 - z_2 } \big). }\eqno\eq  $$
In the last line of each example,
we used the fact that the $\beta \gamma$ propagator
${1 \over z-w}$ satisfies
$    { 1 \over t_1-t_2 } \, { 1 \over t_2-t_3 }
   ={ 1 \over t_1-t_3 } \,
   \{ { 1 \over t_2-t_3 }-{ 1 \over t_2-t_1 } \}.$

{\bf \chapter{Integral Formulas from the Wakimoto Realization}}

 Now we present the systematic derivation of
the integral formulas [\aty].
We find that $\omega$ can be calculated
without using the explicit form for $S_{\alpha}(t)$ and $P(z)$.

\vskip 2mm
\noindent{\bf \S$\,$5.1.}~
 At first we evaluate the correlation
in the case of $P_\lambda(z)$ being an arbitrary polynomial in
${\bf C}[\gamma^\alpha(z)]$.
All we need are the following $S_{\alpha}(t)$, $P(z)$ OPE relations.

 Since $S_{\alpha}(t)$ is constructed
as the left action of $e_{\alpha}$,
the $\beta\gamma$ parts of the screening currents
and the vertex operator satisfy
$$\eqalign {S_\alpha(t_1) S_\beta(t_2)&=\sum_{\gamma  > 0 }
{1 \over {t_1 -t_2}}
  f_{\alpha \beta}^\gamma S_\gamma(t_2)+\cdots, \cr
S_\alpha(t) P(z) &={1 \over {t-z}} (S_\alpha P)(z)+\cdots,}\eqno \eq$$
where $f_{\alpha \beta}^\gamma$ are the structure constants of
${\bf n_+}$
and the operator $(S_\alpha P)(z)\in {\bf C}[ \gamma^\alpha(z) ]$
corresponds to the polynomial
$ (S_\alpha P) = \sum_\beta S_\alpha^\beta (x)
{ \partial \over \partial x^\beta} P(x)
\in {\bf C}[ x^\alpha ]$ .

 From this singularity and the transformation property,
we obtain the {\it ``screening currents Ward identity''};
$$\eqalign{
\langle S_\alpha(&t)  S_{\alpha_1}(t_1) \cdots S_{\alpha_m}(t_m)
 P_1(z_1) \cdots P_n(z_n)\rangle  \cr
\noalign{\vskip 2mm}
&=\sum_{i=1}^m {1 \over {t-t_i}} f_{\alpha \alpha_i}^\beta
 \langle S_{\alpha_1}(t_1) \cdots S_\beta(t_i) \cdots S_{\alpha_m}(t_m)
 P_1(z_1) \cdots P_n(z_n)\rangle  \cr
&+\sum_{a=1}^n {1 \over {t-z_a}}
\langle S_{\alpha_1}(t_1) \cdots S_{\alpha_m}(t_m)
 P_1(z_1) \cdots S_\alpha P_a(z_a) \cdots P_n(z_n)\rangle .}\eqno \eq$$

 This equation gives

\noindent{\bf  THEOREM I.}
 {\it$\,$ For an arbitrary $P_a(z_a)$,
the $\beta \gamma$ correlation $\omega$ is given by
$$ \omega = \sum_{part} \prod_{a=1}^n \langle  \prod_{i \in I_a}
S_{\alpha_i}(t_i)P_a(z_a)\rangle ,\eqno \eq$$
$$ \langle S_{\alpha_1}(t_1) \cdots S_{\alpha_m}(t_m) P(z)\rangle
 =\sum_{perm} {
 \langle (S_{\alpha_1} \cdots S_{\alpha_m} P)(z)\rangle
  \over (t_1-t_2)(t_2-t_3) \cdots (t_m-z)} ,\eqno \eq$$
where $\sum_{part}$ stands for the summation over all the partition of
$I=\{1,2, \cdots ,m\}$ into $n$ disjoint union $I_1 \cup I_2 \cup
\cdots \cup I_n$ and $\sum_{perm}$ the summation over all the
permutation of the elements of $\{1,2, \cdots ,m\}$.}

 The proof is given in  Appendix C.
Theorem I corresponds to the 1st solution of
Schechtman and Varchenko [\sv].

\endpage

\noindent{\bf \S$\,$5.2.}~
 Next we derive the correlation
in the case where $P$ is given by (3.11),
$ P_\lambda^{\{\beta_1,\cdots,\beta_n\}} (z)  $.
A key formula is the $s_\mu(t)$ $F_\beta(z)$ OPE relation (3.9),
which gives a more simple proof of the previous result [\aty].

By definition (3.11), $\beta$ insertion for
$P_\lambda^{\{\beta_1,\cdots,\beta_n\}} (z) $ is given by
$$e^{\lambda \cdot \phi(z)}
P_\lambda^{\{\beta,\beta_1,\cdots,\beta_n\}} (z)
=
\int {dw \over 2\pi i}  F_{\beta}(w)
e^{\lambda \cdot \phi(z)}
P_\lambda^{\{\beta_1,\cdots,\beta_n\}} (z) \eqno\eq$$
Using the $s_\alpha(t)$ $F_{\beta}(z)$ OPE relation (3.9),
we get the following inductive formula
$$\eqalign{\int &\prod_{i=1}^{m}dt_i \langle \prod_{i=1}^{m}
e^{-\alpha_{i} \cdot \phi(t_i)}
e^{\lambda \cdot \phi(z)} \rangle
\langle \prod_{i=1}^{m} S_{\alpha_i}(t_i)
P_\lambda^{\{\beta,\beta_1,\cdots,\beta_n\}} (z) \rangle  \cr
&=\int \prod_{i=1}^{m}dt_i
\sum_{k=1}^m \kappa \delta_{\beta}^{\alpha_k} {2\over \alpha_k^2}
{\partial \over \partial t_k} \langle \prod_{i=1}^{m}
e^{-\alpha_{i} \cdot \phi(t_i)}e^{\lambda \cdot \phi(z)} \rangle
\langle \prod_{i \neq k}^m S_{\alpha_i}(t_i)
P_\lambda^{\{\beta_1,\cdots,\beta_n\}} (z) \rangle.  }\eqno \eq$$
{}From (4.3), we can calculate this derivative
$${\partial \over \partial t_k} \langle \prod_{i=1}^{m}
e^{-\alpha_{i} \cdot \phi(t_i)}e^{\lambda \cdot \phi(z)}  \rangle
={1\over \kappa} (\sum_{l \neq k}^m
  {\alpha_{k} \cdot \alpha_{l} \over t_k-t_l}
-{\alpha_{k} \cdot \lambda        \over t_k-z  })
\langle \prod_{i=1}^{m} e^{-\alpha_{i} \cdot \phi(t_i)}
e^{\lambda         \cdot \phi(z)}  \rangle.\eqno\eq$$
Then we obtain the following
{\it ``currents Ward identity''} including screening currents
$$\eqalign{\langle  \prod_{i=1}^m &S_{\alpha_i} (t_i)
 P_\lambda^{\{\beta,\beta_1,\cdots,\beta_n\}}(z) \rangle  \cr
&=\sum_{k=1}^m \delta_{\beta}^{\alpha_k} {2\over \alpha_k^2}
 (\sum_{l \neq k}^m{\alpha_{k} \cdot \alpha_{l} \over t_k-t_l}
                  -{\alpha_{k} \cdot \lambda \over t_k-z})
\langle  \prod_{i \neq k}^m S_{\alpha_i} (t_i)
P_\lambda^{\{\beta_1,\cdots,\beta_n\}} (z) \rangle .}\eqno \eq $$

 Iterative use of this equation gives the expression for $\omega$.
Note that (5.4) vanishes unless
$\sum_{i=1}^m \alpha_{i} =\sum_{j=1}^n \beta_{j}$.
Then we have

\noindent{\bf  THEOREM II.}
 {\it$\,$ For the basis vectors $P_{\lambda_a}^{I_a}(z_a)$,
the $\beta \gamma$ correlation $\omega$ is given by
$$\omega = \sigma \prod_{a=1}^n        \langle  \prod_{\alpha_i \in I_a}
S_{\alpha_i}(t_i) P_{\lambda_a}^{I_a}(z_a)\rangle,\eqno\eq$$
$$\langle  \prod_{i=1}^n S_{\alpha_i} (t_i)
P_\lambda^{\{\alpha_1,\cdots,\alpha_n\}} (z) \rangle
 =\sigma \prod_{k=1}^n  {2\over \alpha_k^2}
 (\sum_{l=k+1}^n{\alpha_{k} \cdot \alpha_{l} \over t_k-t_l}
                     -{\alpha_{\alpha_k} \cdot \lambda \over t_k-z}) ,
\eqno\eq$$
where $\sigma$ is the symmetrization of the $t$'s associated with the same
$\alpha_i$'s. }
\footnote\dag{
However we must not overcount $\sigma$ in (5.9) and (5.10).}

 The proof is given in Appendix C.
Theorem II coincides with the 2nd solution of
Schechtman and Varchenko [\sv].
 Note that $\langle  \prod_{i=1}^n S_{\alpha_i} (t_i)
P_\lambda^{\{\alpha_1,\cdots,\alpha_n\}} (z) \rangle  $
depends on the ordering of the $\alpha_i$'s of
$P_\lambda^{\{\alpha_1,\cdots,\alpha_n\}}$,
although it is symmetric with respect to $S_{\alpha_i} (t_i)$'s.

\vskip 2mm
\noindent{\bf \S$\,$5.3.}~
 Finally we give the condition for
 the contour of this correlation function.

 The correlation function ( a solution of the KZ equation )
should satisfy the Ward identity for Kac-Moody current $J^a(z)$
such that
$$\langle J^a(z) \prod_{i=1}^n V_{\lambda_i}(w_i) \rangle
= \sum_{i=1}^n { (J^a)_{\lambda_i} \over z-w_i }
\langle \prod_{i=1}^n V_{\lambda_i}(w_i) \rangle .\eqno\eq$$
where $(J^a)_{\lambda_i}$ acts on the primary field
$V_{\lambda_i}(w_i)$.

 From (3.9),
 $E_\alpha(z)$ and $H_i(z)$ insertion satisfies this relation,
but there are the following corrections for $ F_\alpha(z) $
$$\eqalign{&\int\prod_{i=1}^{m}dt_i \langle F_\alpha(z)
\prod_{i=1}^{m} e^{-\alpha_{i} \cdot \phi(t_i)} S_{\alpha_i}(t_i)
\prod_{a=1}^{n} e^{\lambda_a \cdot \phi(z_a)} P_a(z_a) \rangle \cr
&= - \int \prod_{i=1}^{m} dt_i
\sum_{j=1}^m  \kappa \delta_\alpha^{\alpha_j} {2\over \alpha_j^2}
{\partial \over \partial t_j} \langle { 1 \over z-t_j}
\prod_{i=1}^{m} e^{-\alpha_{i} \cdot \phi(t_i)}
\prod_{i\neq j}^{m} S_{\alpha_i}(t_i)
\prod_{a=1}^{n}e^{\lambda_a \cdot \phi(z_a)} P_a(z_a)\rangle \cr
&+ \sum_{b=1}^n { (F_\alpha)_{\lambda_b} \over z-z_b }
\int\prod_{i=1}^{m}dt_i \langle
\prod_{i=1}^{m} e^{-\alpha_{i} \cdot \phi(t_i)} S_{\alpha_i}(t_i)
\prod_{a=1}^{n} e^{\lambda_a \cdot \phi(z_a)} P_a(z_a) \rangle . \cr
}\eqno \eq$$

The paths should be chosen so that the total divergence terms vanish.
For instance we can use the Dotsenko Fateev or Felder type contours
[\df, \fe].

{\bf \chapter{Conclusion and Discussion }}

 In this paper, we have explicitly derived
the correlation function of the chiral primary field
for the WZNW models, based on the Wakimoto realization.
In general it is difficult to calculate this correlator
by simply using the $\beta \gamma$ OPE relation (4.5)
except for some particular cases.
But we made it possible by finding two important relations,
the {\it ``screening currents Ward identity''} (5.2)
and the {\it ``currents Ward identity''}
including the screening currents (5.8).

In addition, we wish to make two further comments.

 i).~ In terms of the differential representation of the KZ equation,
where the matrix $\Omega_{ab}$ is replaced by
some differential operators,our result gives only the solution
over polynomial ${\bf C}[x^{\alpha}]$.
But in the case where the level of the KM algebra is non integer,
it is natural that a non polynomial solution appears.
So the structure of the non integer level solution should be analyzed.
The dimension of the space of these solutions ( i.e. fusion rules )
is now under investigation [\ay].

 ii).~ In the differential realization of the simple Lie algebras,
besides the ``right action'' (2.2), there is also the ``left action''
$ \widehat J({\partial \over \partial x},x,\lambda) $ defined by
$$\widehat  J({\partial \over \partial x},x,\lambda)
      \, Z \, \vert \widehat \lambda \rangle
= - J \, Z \, \vert \widehat \lambda \rangle  .\eqno\eq  $$
where  $Z \equiv \exp ( \sum_{\alpha} x^\alpha e_\alpha ) $,
and $\vert \widehat \lambda \rangle $
is the lowest weight vector such that
$f_{\alpha} \vert \widehat \lambda \rangle = 0$,
$h_i        \vert \widehat \lambda \rangle =
  \lambda_i \vert \widehat \lambda \rangle $.
 As mention in Appendix A.2,
        $  J({\partial \over \partial x},x,\lambda) $ and
$\widehat  J({\partial \over \partial x},x,\lambda) $
are related by replacing $ x^\alpha $ with $ - x^\alpha $ and
$ \lambda_a $ with $ -\lambda_a $.
Note that one of these operators is (2.5).

 Similarly, there is another Wakimoto realization in addition to (3.5),
which is given by replacing $ \beta_\alpha $ with $ - \beta_\alpha $,
$ \gamma^\alpha $ with $ - \gamma^\alpha $ and
$ \phi_i $ with $ - \phi_i $.
 Since one of these left currents is the screening current (3.8),
they would be related to the quantum groups [\qg, \sva].

{\bf \ack}

 This work has been carried out in collaboration with Y. Yamada.
I am grateful to him for help in many ways.
 I would like to thank  A. Tsuchiya
and the members of KEK theory group  for valuable discussions.
I would also like to thank to N. A. McDougall
for a careful reading of the manuscript.

{\bf \Appendix{A}}

 In this Appendix we present
some useful formulas for the free fields realization,
\footnote{\dag} {Our parametrization for $Z$ is the same as [\iko].
Restricted to the $sl(n)$ case,
the parametrization of [\ffr] takes a more simple form.}
and the proof of (2.6).

\vskip 2mm
\noindent{\bf \S$\,$A.1.}~ We give some helpful relations.

\noindent{\bf PROPOSITION.}
 {\it $\,$  Gauss decomposition of the right action is given by
the following differential operators }
$$\eqalign{Z \; \exp( t e_\alpha ) &=
\exp( t E_\alpha ({\partial\over\partial x},x) + O(t^2) )  \; Z, \cr
Z \; \exp( t h_i )
&= \exp(t h_i )
\exp( t \widehat H_i({\partial\over\partial x},x) + O(t^2) ) \; Z, \cr
Z \; \exp( t f_{\alpha_i} )
&= \exp( t f_{\alpha_i} ) \exp(t h_i x^{\alpha_i} )
\exp( t \widehat F_{\alpha_i}({\partial\over\partial x},x)  + O(t^2) )
\;Z,}\eqno\eq$$
{\it where }
$$\eqalign{ E_\alpha({\partial\over\partial x},x)
&= \sum_{\beta }  \big( \delta_\alpha^\beta
-{1\over2}\sum_{\gamma>0} f_{\alpha \gamma}^\beta x^\gamma
+ O(x^2) \big)   {\partial \over \partial x^\beta}, \cr
 \widehat{H}_i({\partial\over\partial x},x)
&= - \sum_{\beta >0}   (\nu_i,\beta)x^\beta
   {\partial \over \partial x^\beta} , \cr
 \widehat{F}_{\alpha_i}({\partial\over\partial x},x)
&= - \sum_{\beta }
 \big( \sum_{\gamma } f_{-\alpha_i \gamma}^\beta x^\gamma
+ {1\over2} (\nu_i,\beta) x^{\alpha_i} x^\beta + O(x^3) \big)
   {\partial \over \partial x^\beta}  .}\eqno\eq$$

\noindent \underbar { Proof.  }~
 It is given by using the Baker-Campbell-Hausdorff formula
$$\eqalign{e^X e^Y &=
\exp \big( X+Y+{1\over 2}[X,Y]+{1\over 12} ( [X,[X,Y]] + [Y,[Y,X]] )
+ \cdots \big),\cr
e^{-Y} e^X e^Y &= \exp \big( X + [X,Y] + O(Y^2) \big).}\eqno\eq$$
The right action of $e_\alpha \in {\bf n_+} $ is
$$\eqalign{Z \; \exp ( t e_\alpha )
&=\exp\big( \sum_\beta x^\beta e_\beta  + t \big( e_\alpha
        - {1\over2}\sum_{\beta\gamma} f_{\alpha\beta}^\gamma x^\beta
		e_\gamma + O(x^2) \big)			+ O(t^2) \big),\cr
&=\exp\big( \sum_\beta x^\beta e_\beta
        + t \big( {\partial \over \partial x^\alpha}
        - {1\over2} \sum_{\beta\gamma} f_{\alpha\beta}^\gamma x^\beta
		  {\partial \over \partial x^\gamma} + O(x^2) \big)
		  \sum_\beta x^\beta e_\beta 		+ O(t^2) \big) \cr
&=\exp\big( t \big( {\partial \over \partial x^\alpha}
        - {1\over2} \sum_{\beta\gamma} f_{\alpha \beta}^\gamma x^\beta
		  {\partial \over \partial x^\gamma} +O(x^2)  \big)
		  	+ O(t^2) \big) \; Z. \cr}\eqno\eq$$
For $h_i \in {\bf h} $, Gauss decomposition of the right action is
$$\eqalign{Z \; \exp ( t h_i ) &=\exp    ( t h_i )
  \exp\big( \sum_\beta x^\beta e_\beta
            - t \sum_\beta (\nu_i \cdot \beta ) x^\beta e_\beta
			+ O(t^2) \big) \cr
&=\exp    ( t h_i )
  \exp\big( - t \sum_\beta (\nu_i \cdot \beta ) x^\beta
			{\partial \over \partial x^\beta}
			+ O(t^2)  \big) \; Z. \cr}\eqno\eq$$
For $f_{\alpha_i} \in {\bf n_-}$, with simple root $\alpha_i$
$$\eqalign{Z \; \exp ( t f_{\alpha_i} ) &=\exp    ( t f_{\alpha_i} )
  \exp\big( \sum_\beta x^\beta e_\beta     + t h_i x^{\alpha_i}
			- t \sum_{\beta\gamma}
			f_{-\alpha_i \beta }^\gamma x^\beta e_\gamma
			+ O(t^2) \big) \cr
&=\exp    ( t f_{\alpha_i} )
  \exp    ( t h_i x^{\alpha_i}  ) \; \Theta, \cr}\eqno\eq$$
$$\eqalign{\Theta \;&=\exp\big( \sum_\beta x^\beta e_\beta
	-t  \big( \sum_{\beta\gamma}
		f_{-\alpha_i \beta }^\gamma x^\beta e_\gamma
			+{1 \over 2}  \sum_\beta
			(\nu_i \cdot \beta) x^{\alpha_i}  x^\beta e_\beta
			+  O(x^3) \big) + O(t^2) \big) \cr
&=\exp\big( -t \big( 			 \sum_{\beta\gamma}
         	f_{-\alpha_i \beta }^\gamma x^\beta
			{\partial \over \partial x^\gamma}
			+{1 \over 2} \sum_\beta
			(\nu_i \cdot \beta) x^{\alpha_i}  x^\beta
			{\partial \over \partial x^\beta} + O(x^3) \big)
		    + O(t^2) \big) \; Z. \cr}\eqno\eq$$\hfill Q.E.D.

Moreover,
if we act with the highest weight vector $\langle \lambda \vert $
from the left in (A.1)
and take the infinitesimal limit with respect to $t$,then we get
the differential operators $E_{\alpha}$, $F_{\alpha_i}$ and $H_i$
corresponding to the Chevalley generators
$e_{\alpha} $, $f_{\alpha_i} $ and $h_i $
$$\eqalign{E_\alpha({\partial\over\partial x},x,\lambda)
&=E_\alpha({\partial\over\partial x},x) , \cr
H_i({\partial\over\partial x},x,\lambda)
&= \widehat{H}_i({\partial\over\partial x},x)   + \lambda_i, \cr
F_{\alpha_i}({\partial\over\partial x},x,\lambda)
&= \widehat{F}_{\alpha_i}({\partial\over\partial x},x)
+ \lambda_i x^{\alpha_i} .}\eqno\eq$$

Finally, for the left action in (2.4) and (6.1),
the Gauss decompositions ( of inverse direction )  are given by
taking the inverse of the right action and
reversing the signs of $x^\alpha $'s as follows
$$\eqalign{\exp( -t e_\alpha ) \; Z&=
\exp( t E_\alpha (-{\partial\over\partial x},-x) + O(t^2) )  \; Z, \cr
\exp( -t h_i )\; Z
&= \exp( t \widehat H_i(-{\partial\over\partial x},-x) + O(t^2) )
\; Z \; \exp(-t h_i ),\cr
\exp(- t f_{\alpha_i} ) \;Z
&=
\exp( t \widehat F_{\alpha_i}(-{\partial\over\partial x},-x)+ O(t^2) )
\; Z \; \exp(t h_i x^{\alpha_i} )\exp(- t f_{\alpha_i} ) .}\eqno\eq$$
For instance
$$S_\alpha({\partial\over\partial x},x)
=E_\alpha(-{\partial\over\partial x},-x).\eqno\eq$$


\vskip 2mm
\noindent{\bf \S$\,$A.2.}~Next we derive (2.6),
by using the associativity for the Gauss decomposition.

\noindent \underbar { Proof of (2.6.a). }~
$$\eqalign{\big( \exp(-s e_\beta) \;Z \; \big)
\exp( t e_\alpha )
&= \exp( s S_\beta + O(s^2) ) \exp( t E_\alpha + O(t^2) ) \; Z, \cr
\exp(-s e_\beta) \big( \;Z \; \exp( t e_\alpha ) \big)
&= \exp( t E_\alpha + O(t^2) ) \exp( s S_\beta + O(s^2) ) \;Z.}\eqno\eq$$
{}From the associativity, $E_\alpha$ and $S_\alpha$ commute.\hfill Q.E.D.

\noindent \underbar { Proof of (2.6.b). }~
$$\big( \exp(-s e_\alpha) \;Z \; \big) \exp( t h_i )
=\exp( s S_\alpha + O(s^2) )
\exp(t ( h_i +  \widehat H_i ) + O(t^2) ) \;Z,\eqno\eq$$
$$\eqalign{\exp(-s e_\alpha) \big( \;Z \; \exp( t h_i ) \big)
= &\exp(t ( h_i + \widehat H_i ) + O(t^2) )
\exp( s S_\alpha + O(s^2) ) \cr
&\exp(- st (\nu_i \cdot \alpha)  S_\alpha + O(s^2)+O(t^2) ) \; Z ,
}\eqno\eq$$
where (A.13) comes from
$$\exp(-s e_\alpha) \exp(t h_i )
=\exp(th_i)\exp( st (\nu_i \cdot \alpha) e_\alpha + O(s^2)+O(t^2) )
\exp(-s e_\alpha).\eqno\eq$$
If we act with the highest weight vector
$\langle \lambda \vert $ from the left, then
$$\eqalign{&\exp( s S_\alpha + O(s^2) )
\exp( t H_i + O(t^2) ) \langle \lambda \vert \;Z \cr
=&\exp( t H_i + O(t^2) ) \exp( s S_\alpha + O(s^2) )
\exp(- st (\nu_i \cdot \alpha)  S_\alpha + O(s^2)+O(t^2) )
\langle \lambda \vert \;Z.}\eqno\eq$$\hfill Q.E.D.

\noindent \underbar { Proof of (2.6.c). }~
$$\eqalign{\big( \exp(-s e_{\alpha_j}) \;Z \; \big)
&\exp( t f_{\alpha_i} )\cr
=&\exp( s S_{\alpha_j} + O(s^2) )
\exp( t ( f_{\alpha_i} + h_i x^{\alpha_i} +
 \widehat F_{\alpha_i}) + O(t^2) ) \;Z,\cr}\eqno\eq$$
$$\eqalign{\exp(-s e_{\alpha_j}) \big( \;Z \;
&\exp( t f_{\alpha_i} ) \big)\cr
=& \exp( t ( f_{\alpha_i} + h_i x^{\alpha_i} +
 \widehat F_{\alpha_i} )+ O(t^2) )
\exp( s S_{\alpha_j} + O(s^2) ) \cr
&\exp(-st( \delta_{ij} h_i
+ (\nu_i \cdot \alpha_j) x^{\alpha_i} S_{\alpha_j} )+O(s^2)+O(t^2))
\;Z, }\eqno\eq$$
where (A.17) comes from
$$\eqalign{&\exp(-s e_{\alpha_j})
\exp( t (f_{\alpha_i} + h_i x^{\alpha_i}) )\cr
=&\exp( t (f_{\alpha_i} + h_i x^{\alpha_i}) )
\exp(st (- \delta_{ij} h_i
+  (\nu_i \cdot \alpha_j) e_{\alpha_j} x^{\alpha_i} )+O(s^2)+O(t^2))
\exp(-s e_{\alpha_j}) .}\eqno\eq$$
Then we get
$$\eqalign{&\exp( s S_{\alpha_j} + O(s^2) )
\exp( t F_{\alpha_i} + O(t^2) ) \langle \lambda \vert \;Z \cr
=&\exp( t F_{\alpha_i} + O(t^2) )
\exp( s S_{\alpha_j} + O(s^2) ) \cr
&\exp(-st (\delta_{ij} \lambda_i
+(\nu_i \cdot \alpha_j) x^{\alpha_i} S_{\alpha_j} )+O(s^2)+O(t^2))
\langle \lambda \vert \;Z.}\eqno\eq$$\hfill Q.E.D.

{\bf \Appendix{B}}

 In this Appendix we explain another derivation of
the ``current Ward identity'' including screening currents (5.8).
The outline of this derivation is indicated in [\aty].
Although it is more complicated than that in the text,
it shows the clear relation between the 1st and 2nd solution.

\vskip 2mm
\noindent{\bf \S$\,$B.1.}~ We start with evaluating
$\langle (S_{\alpha_1} \cdots S_{\alpha_m} P)(z)\rangle $ in (5.4),
in the case where $P$ is given in (3.11).
We will show that this correlation can be calculated
by using only finite dimensional algebra.

 Note that only the c-number term of the operator
$(S_{\alpha_1} \cdots S_{\alpha_m} P)(z) \in {\bf C}[\gamma^\alpha (z)]$ has a
nonzero contribution to the correlation
$\langle (S_{\alpha_1} \cdots S_{\alpha_m} P)(z)\rangle $.
 Moreover, the constant term of the operator
$(S_{\alpha_1} \cdots S_{\alpha_m} P)(z) \in {\bf C}[\gamma^\alpha (z)]$ is the
same as that of the polynomial
$S_{\alpha_1} \cdots S_{\alpha_m} P \in {\bf C}[x^\alpha]$.

 By definition of the differential representation (2.4) and (2.10),
$$S_{\alpha_1} \cdots S_{\alpha_m} P  =
(-1)^m \langle \lambda \vert e_{\alpha_m} \cdots e_{\alpha_1}
  \, Z \, f_{\beta_1} \cdots f_{\beta_n} \vert \lambda \rangle.\eqno\eq$$
Since the constant term of this polynomial is given by the value at
$x^\alpha=0$,  i.e. $Z=1$ , the correlator
$\langle (S_{\alpha_1} \cdots S_{\alpha_m} P)(z)\rangle $
is nothing but the ``Shapovalov form'' $(\S 2.1)$ up to a sign,
$$\langle (S_{\alpha_1} \cdots S_{\alpha_m} P)(z)\rangle =
(-1)^m \langle \lambda \vert e_{\alpha_m} \cdots e_{\alpha_1}
f_{\beta_1} \cdots f_{\beta_n} \vert \lambda \rangle . \eqno\eq$$

The Shapovalov form enjoys the inductive formula
$$\langle \lambda  \vert \prod_{i=m}^1 e_{\alpha_i}
        f_{\beta} \prod_{j=1}^n f_{\beta_j} \vert \lambda  \rangle
=-\sum_{k=1}^m \delta_{\beta}^{\alpha_k}{2\over \alpha_k^2}
(\sum_{l=k}^m \alpha_{k} \cdot \alpha_{l}
              - \alpha_{k} \cdot \lambda )
\langle \lambda \vert \prod_{i \neq k}^1 e_{\alpha_i}
        \prod_{j=1}^n      f_{\beta_j} \vert \lambda \rangle .\eqno \eq$$
So we have the following property
$$\langle \lambda \vert
e_{\alpha_m} \cdots [ e_{\alpha_i},e_\alpha ] \cdots e_{\alpha_1}
f_\beta \prod_{j=1}^n f_{\beta_j} \vert \lambda \rangle
=-\delta_\beta^\alpha {2\over \alpha^2} (\alpha \cdot \alpha_{i})
\langle \lambda \vert e_{\alpha_m} \cdots e_{\alpha_1}
\prod_{j=1}^n f_{\beta_j} \vert \lambda \rangle + \cdots, \eqno\eq$$
$$\langle \lambda \vert e_\alpha e_{\alpha_m} \cdots e_{\alpha_1}
f_\beta \prod_{j=1}^n f_{\beta_j} \vert \lambda \rangle
= \delta_\beta^\alpha {2\over \alpha^2} (\alpha \cdot \lambda)
\langle \lambda \vert e_{\alpha_m} \cdots e_{\alpha_1}
\prod_{j=1}^n f_{\beta_j} \vert \lambda \rangle + \cdots. \eqno\eq$$
Here, and in the following,
$+ \cdots $ means a sum of terms proportional to
$\delta_\beta^{\alpha_k}$ for $k=1,\cdots ,q$.

 Combining this Shapovalov form and Theorem I, we obtain
$$\eqalign{\langle
S_{\alpha_1}(&t_1) \cdots [S_{\alpha},S_{\alpha_i}](t_i) \cdots
S_{\alpha_n}(t_n)
P_\lambda^{\{\beta,\beta_1,\cdots,\beta_n\}}(z) \rangle  \cr
&=\delta^{\alpha}_{\beta} {2\over \alpha^2} (\alpha \cdot \alpha_{i})
\langle  \prod_{i =1}^m S_{\alpha_i} (t_i)
P_\lambda^{\{\beta_1,\cdots,\beta_n\}} (z) \rangle + \cdots, \cr
}\eqno\eq$$
$$\eqalign{\langle S_{\alpha_1}(&t_1) \cdots S_{\alpha_n}(t_n)
\big( S_{\alpha} P_\lambda^{\{\beta,\beta_1,\cdots,\beta_n\}}\big)(z)\rangle
\cr
&=- \delta^{\alpha}_{\beta} {2\over \alpha^2} (\alpha_{1} \cdot \lambda
)\langle  \prod_{i =1}^m S_{\alpha_i} (t_i)
P_\lambda^{\{\beta_1,\cdots,\beta_n\}} (z) \rangle + \cdots ,}\eqno\eq$$
where $[S_{\alpha},S_{\beta}](t)$
means $\sum_\gamma f_{\alpha\beta}^\gamma S_\gamma(t)$.

\endpage
\vskip 2mm
\noindent{\bf \S$\,$B.2.}~\underbar{Proof of (5.8). }~

 From (5.4) and (B.3), we can expand the left hand side
by $\delta^{\alpha_k}_{\beta}$  as follows
$$\langle  \prod_{i=1}^m S_{\alpha_i} (t_i)
P_\lambda^{\{\beta,\beta_1,\cdots,\beta_n\}}(z) \rangle
= \sum_{k=1}^m \delta^{\alpha_k}_{\beta} \Theta_k (t,z,\alpha,\beta).
\eqno\eq$$
Since ${\{\alpha_1,\cdots,\alpha_n\}}$ is arbitrary,
the $ \delta^{\alpha_k}_{\beta} $'s
constitute a linearly independent basis.
So the function $\Theta_k (t,z,\alpha,\beta)$
is uniquely defined as a coefficient.

 To obtain the $\Theta_1 (t,z,\alpha,\beta)$, we use (5.2). Then,
$$\eqalign{(B.8)&=\sum_{i=2}^m {1 \over {t_1-t_i}} \langle
S_{\alpha_2}(t_2) \cdots [S_{\alpha_1},S_{\alpha_i}](t_i) \cdots
S_{\alpha_n}(t_n)
P_\lambda^{\{\beta,\beta_1,\cdots,\beta_n\}}(z) \rangle  \cr
&+{1 \over {t_1-z}}
\langle S_{\alpha_2}(t_2) \cdots S_{\alpha_n}(t_n)
\big( S_{\alpha_1}
P_\lambda^{\{\beta,\beta_1,\cdots,\beta_n\}}\big)(z)\rangle.}\eqno\eq$$
 From (B.6) and (B.7), we have
$$(B.8)= \delta^{\alpha_1}_{\beta} {2\over \alpha_1^2}
  (\sum_{l=2}^m{\alpha_{1} \cdot \alpha_{l} \over t_1-t_l}
              -{\alpha_{1} \cdot \lambda        \over t_1-z}  )
\langle  \prod_{i =2}^m S_{\alpha_i} (t_i)
P_\lambda^{\{\beta_1,\cdots,\beta_n\}} (z) \rangle
+ \cdots .\eqno\eq$$
The 1st term of this gives the $\Theta_1 (t,z,\alpha,\beta)$.

 Since (B.8) is symmetric with respect to
$ S_{\alpha_i} (t_i) $'s, we get
$$\Theta_k (t,z,\alpha,\beta)={2\over \alpha_k^2}
(\sum_{l \neq k}^m{\alpha_{k} \cdot \alpha_{l} \over t_k-t_l}
                  -{\alpha_{k} \cdot \lambda        \over t_k-z})
\langle  \prod_{i \neq k}^m S_{\alpha_i} (t_i)
P_\lambda^{\{\beta_1,\cdots,\beta_n\}} (z) \rangle .\eqno\eq$$\hfill Q.E.D.

{\bf \Appendix{C}}

 In this Appendix we give the proofs of Theorem I and Theorem II.

\vskip 2mm
\noindent{\bf \S$\,$C.1.}~Proof of Theorem I from (5.2).

\noindent \underbar{Proof of (5.4).}~
 We prove (5.4) by induction on the number of $S_\alpha(t)$,
which we denote by  $n$.   For $n=1$, the assertion is valid.
Assume that (5.4) holds for all $n\leq m$.It is sufficient to show
$$\eqalign{
\langle S_{\alpha}&(t)S_{\alpha_1}(t_1) \cdots S_{\alpha_m}(t_m) P(z)\rangle
\cr
&=\sum_{perm}
{\langle (S_{\alpha}S_{\alpha_1} \cdots S_{\alpha_m} P)(z)\rangle
\over (t-t_1)(t_1-t_2) \cdots (t_m-z)}. }\eqno \eq$$
Both sides are global 1-form on the sphere
with respect to $t$.
On the right hand side, the residue at the pole $z$ is
$$\eqalign{{1\over 2\pi i }\int_z dt
&\sum_{perm}{ \langle (S_{\alpha_1} \cdots S_{\alpha_m} S_{\alpha} P)(z)\rangle
 \over (t_1-t_2)(t_2-t_3) \cdots (t_m-t)(t-z)} \cr
&=\sum_{perm}{\langle (S_{\alpha_1} \cdots S_{\alpha_m} S_{\alpha} P)(z)\rangle
 \over (t_1-t_2)(t_2-t_3) \cdots (t_m-z)}.}\eqno\eq$$
And the residue at the pole $t_i$ is
$$\eqalign{{1\over 2\pi i }\int_{t_i} dt &\sum_{perm} \big(
{\langle (S_{\alpha_1}\cdots S_{\alpha}S_{\alpha_i}\cdots S_{\alpha_m}
P)(z)\rangle\over (t_1-t_2) \cdots (t_{i-1}-t)(t-t_i)(t_i-t_{i+1}) \cdots
(t_m-z)}\cr
&\; \hskip10mm
+{\langle(S_{\alpha_1}\cdots S_{\alpha_i}S_{\alpha}\cdots S_{\alpha_m}
P)(z)\rangle
\over (t_1-t_2) \cdots (t_{i-1}-t_i)(t_i-t)(t-t_{i+1}) \cdots (t_m-z)}
\big) \cr
&= \sum_{perm} {\langle (S_{\alpha_1}\cdots [S_{\alpha},S_{\alpha_i}]
\cdots S_{\alpha_m} P)(z) \rangle
\over (t_1-t_2) \cdots (t_{i-1}-t_i) (t_i-t_{i+1}) \cdots (t_m-z)},
}\eqno\eq$$
By the inductive hypothesis,
the poles and the residues are the same on both sides of (C.2),
so they must be equal.\hfill Q.E.D.

\noindent \underbar{ Proof of (5.3).}~
 This is also given by induction on the number of $S_\alpha(t)$
using simple combinatrics.  For $n=1$, (5.3) holds.
Assume that the assertion is valid for all $n\leq m$.
Let  $\sum_{part(m)}$ be the summation over all the partition of
$I=\{1,2, \cdots ,m\}$ into $n$ disjoint union
$I_1^m \cup I_2^m \cup \cdots \cup I_n^m$,
then we get from (5.2)
$$ \eqalign{\langle S_{\alpha_{m+1}}(t_{m+1})
&\prod_{i=1}^{m} S_{\alpha_i}(t_i) \prod_{a=1}^{n}P_a(z_a)\rangle \cr
&=\sum_{part(m)} \sum_{a=1}^n \langle  S_{\alpha_{m+1}}(t_{m+1})
\prod_{i \in I_a^m} S_{\alpha_i}(t_i)P_a(z_a)\rangle
\prod_{b \neq a}^n
\langle  \prod_{i \in I_b^m} S_{\alpha_i}(t_i)P_b(z_b)\rangle \cr
&=\sum_{part(m+1)} \prod_{a=1}^n
\langle  \prod_{i \in I_a^{m+1}} S_{\alpha_i}(t_i)P_a(z_a)\rangle
}\eqno \eq$$\hfill Q.E.D.


\vskip 2mm
\noindent{\bf \S$\,$C.2.}~Proof of Theorem II  from (5.8).

\noindent \underbar{Proof of (5.10).}~
Let $I$ be an ordered set of $\alpha_i$'s.
We can expand the left hand side of (5.10) by
$\delta^I_{ \{ \alpha_1\cdots\alpha_n \} }$ with no ambiguity, as
$$\langle  \prod_{i=1}^n S_{\alpha_i} (t_i)
P_\lambda^{\{\alpha_1,\cdots,\alpha_n\}}(z) \rangle
= \sum_I \delta^I_{ \{ \alpha_1\cdots\alpha_n \} }
\Theta_I (t,z,\alpha),\eqno\eq$$

{}From (5.8),
$\Theta_{ \{ \alpha_1\cdots\alpha_n \} } (t,z,\alpha)$ is easy to find
$$\Theta_{ \{ \alpha_1\cdots\alpha_n \} } (t,z,\alpha)
 = \prod_{k=1}^n  {2\over \alpha_k^2}
 (\sum_{l=k+1}^n{\alpha_{k} \cdot \alpha_{l} \over t_k-t_l}
              -{\alpha_{k} \cdot \lambda \over t_k-z}) . \eqno \eq$$
By the symmetry with respect to $S_{\alpha_i} (t_i) $,
we can get the other $\Theta_I (t,z,\alpha)$ similarly.
Moreover $\sum_I \delta^I_{ \{ \alpha_1\cdots\alpha_n \} } $
is nothing but the symmetrization of the $t$'s
associated with the same $\alpha_i$'s. \hfill Q.E.D.

\noindent \underbar{ Proof of (5.9).}~
It is easy; notice only that
$\sum_{pert}$ is no more than the symmetrization of the  $t$'s
associated with the same $\alpha_i$'s belonging to different $I_a$.
\hfill\break.\hfill Q.E.D.

\par \penalty-400 \vskip\chapterskip
   \spacecheck\referenceminspace \immediate\closeout\referencewrite
   \referenceopenfalse
   \line{\fourteenbf \hfil  References \hfil}\vskip\headskip
   \input reference.aux   

\bye